\documentclass[floatfix,prc,twocolumn,showpacs,nofootinbib]{revtex4}
\usepackage{natbib}
\usepackage{times}
\usepackage{amssymb,amsbsy,amsmath,amsfonts}
\usepackage{graphicx}
\usepackage{float}
\usepackage{morefloats}
\usepackage{rotating}

\begin{document}
\title {Vector meson-vector meson interaction in a hidden gauge unitary approach}

\author{L. S. Geng and E. Oset}
 \affiliation{
Departamento de F\'{\i}sica Te\'orica and IFIC, Universidad de
Valencia-CSIC, E-46071 Valencia, Spain}

\begin{abstract}
The formalism developed recently to study vector meson--vector meson
interaction, and applied to the case of $\rho\rho$, is extended to
study the interaction of the nonet of vector mesons among
themselves. The interaction leads to poles of the scattering matrix
corresponding to bound states or resonances. We show that 11
states (either bound or resonant) get dynamically generated
 in nine strangeness-isospin-spin
channels. Five of them can be identified with those reported in the
PDG, i.e., the $f_0(1370)$, $f_0(1710)$, $f_2(1270)$, $f'_2(1525)$,
and $K^*_2(1430)$. The masses of the latter three tensor states have
been used to fine-tune the free parameters of the unitary approach,
i.e., the subtraction constants in evaluating the vector meson
-vector meson loop functions in the dimensional regularization scheme.
The branching ratios of these five dynamically generated states are
found to be consistent with data. The existence of the other six
states should be taken as predictions to be tested by future
experiments.

\end{abstract}
\pacs{13.75.Lb 	Meson–meson interactions,14.40.Cs Other mesons with S=C=0, mass $<$ 2.5 GeV,
14.40.Ev Other strange mesons, 12.40.Yx Hadron mass models and calculations   }
\date{\today}
 \maketitle

\section{Introduction}
Although Quantum Chromodynamics (QCD) has been generally accepted as
 the underlying theory of the strong interaction, due to the asymptotic freedom, however, its
application at low energies around 1 GeV is highly problematic.
Even in the case of Lattice QCD, one still has to face many problems.
Therefore, one often turns to various effective theories or models. Chiral
symmetry, related with the small masses of  $u$, $d$, $s$ quarks, provides a
general principle for constructing effective field theories to study
low-energy strong-interaction phenomena. In this respect, chiral perturbation theory,
$\chi$PT, has been rather successful in studies of low-energy hadronic
phenomena~\cite{Weinberg:1978kz,Gasser:1984gg,Meissner:1993ah,Bernard:1995dp,Pich:1995bw,Ecker:1994gg}.
See, for instance, Ref.~\cite{Scherer:2002tk} for a pedagogical introduction and
Ref.~\cite{Bernard:2007zu} for recent developments in the one-baryon sector.

However, pure perturbation theory cannot describe the low-lying
resonances. The breakthrough came with the application of unitary
techniques in the conventional chiral perturbation theory, enabling
one to study higher energy regions hitherto inaccessible, while employing
chiral Lagrangians. The
unitary extension of chiral perturbation theory, U$\chi$PT, has been
successfully applied to study meson-baryon and meson-meson
interactions.  Several unitarization approaches have been developed over the years,
including the Inverse Amplitude Method~\cite{Dobado:1996ps,Oller:1998hw}, dispersion relations (the $N/D$ method)~\cite{Oller:1998zr,Oller:2000fj},
or in terms of coupled channel Bethe-Salpeter equation~\cite{Kaiser:1995eg,Oller:1997ti,Kaiser:1998fi}.

So far, the unitary chiral approach has been applied to study
the self-interaction of the octet of pseudoscalars of the $\pi$
\cite{Oller:1997ti,Kaiser:1998fi,Markushin:2000fa,Dobado:1996ps,Oller:1998hw},
which provides the low lying scalar mesons,
 the interaction of the octet of pseudoscalars of the $\pi$  with
 the octet of baryons of the proton, which generates $J^P=1/2^-$ baryonic
 resonances
 \cite{Kaiser:1995eg,Oset:1997it,Oller:2000fj,GarciaRecio:2003ks,Jido:2003cb,GarciaRecio:2005hy,Hyodo:2002pk},
 the interaction of the octet of pseudoscalars of the $\pi$ with the
decuplet of baryons of the $\Delta$
\cite{Kolomeitsev:2003kt,Sarkar:2004jh}, which leads to $J^P=3/2^-$
baryon resonances, and the interaction of the octet of pseudoscalars of
the $\pi$ with the nonet of vector mesons of the $\rho$, which leads to
axial vector meson resonances \cite{Lutz:2003fm,Roca:2005nm}. These
studies sometimes report ``surprising'' results, such as the
existence of two $\Lambda(1405)$ states and  two $K_1(1270)$ states.
Both have found some experimental
support~\cite{Magas:2005vu,Geng:2006yb}. This approach has also been
extended to study systems including a heavy quark, charm or bottom,
the so-called heavy-light
systems~\cite{Kolomeitsev:2003ac,Hofmann:2003je,Guo:2006fu,Gamermann:2006nm},
and to study three-body
resonances~\cite{MartinezTorres:2007sr,MartinezTorres:2008gy}.

The interaction of vector mesons with vector mesons or vector mesons
with baryons has received little attention. One exception is the
work of Ref.~\cite{AlvarezRuso:2002ib} where the vector-vector interaction is used
to provide collision rates of vector mesons in heavy ion collisions.
However,  in a recent work~\cite{Molina:2008jw}, the task of finding
bound states of $\rho\rho$ mesons was undertaken, using unitary
techniques with the interaction vertices derived from the hidden-gauge  Lagrangians~\cite{Bando:1984ej,Bando:1987br}. Using
as input the vertices provided by these Lagrangians and unitarizing
the amplitudes via the Bethe-Salpeter equation, two poles were found
on the complex plane: one in $(I,S)=(0,0)$ and the other in
$(I,S)=(0,2)$ sector, which were identified with the $f_0(1370)$ and
the $f_2(1270)$ states of the PDG~\cite{Amsler:2008zz}. The
formalism provides naturally a stronger attraction for the tensor
channel than for the scalar channel. A study of the radiative decays
of these two states based on this approach has been
performed~\cite{Nagahiro:2008um}.

The main purpose of the present paper is to extend the formalism
developed in Ref.~\cite{Molina:2008jw} to study vector meson-vector
meson interaction in all
 possible strangeness-isospin-spin channels.

This paper is organized as follows: In Sec. II, we write down the
hidden-gauge Lagrangians and briefly describe the several mechanisms
that contribute to tree-level transition amplitudes, including four-vector-contact interaction, $s$, $t$, and $u$-channel vector
exchange, and box diagrams that provide decays to two pseudoscalars.
We also explain in detail the approximations involved to make
calculations feasible and the arguments supporting these
approximations. In Sec. III, we look for poles on the complex plane
and present results channel by channel. We show  results both
without and with the decay mechanism to two pseudoscalars. We also
calculate the residues of these poles, which quantify the couplings
of these states to different coupled channels and play a role in
studies of their radiative decays. Section IV contains a brief
summary and our main conclusions.

\section{Formalism}
In this work, as in Refs.~\cite{Molina:2008jw,Nagahiro:2008um}, we
use the Bethe-Salpeter equation method to unitarize the amplitudes.
In this approach, the unitarized $T$ amplitudes in coupled channels
and $s$ wave can be written as
\begin{equation}
T=V+VGT=(1-VG)^{-1}V,
\end{equation}
where $V$ stands for  the tree-level transition amplitudes, and $G$
is a diagonal matrix with its element the vector meson--vector meson loop
function:
\begin{equation}\label{eq:Gsharp}
G=i\int\frac{d^4q}{(2\pi)^4}\frac{1}{q^2-M^2_{V1}}\frac{1}{q^2-M^2_{V2}},
\end{equation}
where $M_{V1}$ and $M_{V2}$ are the masses of the two vector-mesons.

\begin{figure}[t]
\includegraphics[scale=0.35]{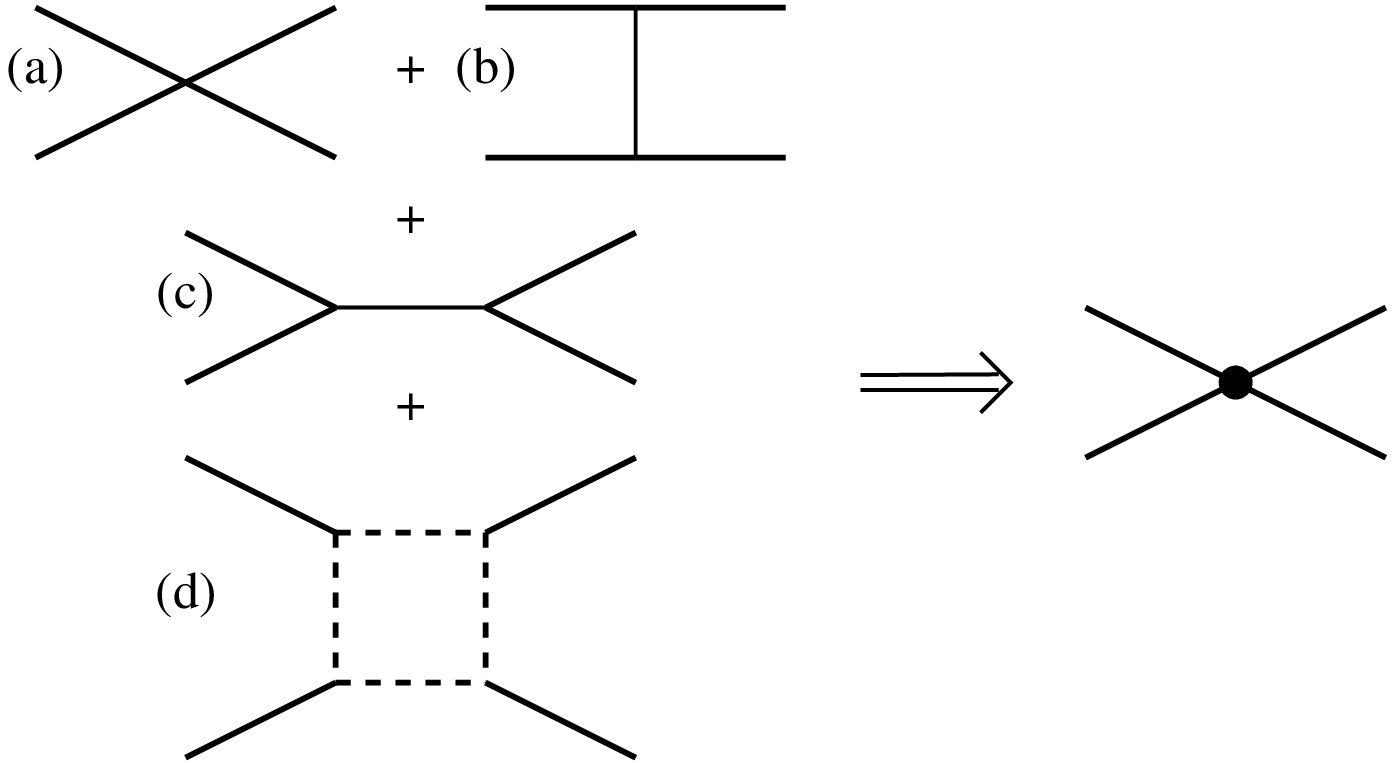}
\caption{The mechanisms contributing to the tree-level vertex of vector-vector scattering,
which appears as $V$ in the coupled channel Bethe-Salpeter equation.\label{fig:dia1}}
\end{figure}

As explained in Ref.~\cite{Molina:2008jw} and also shown in
Fig.~\ref{fig:dia1}, four possible mechanisms contribute to the
tree-level transition amplitudes $V$: (1) four-vector-contact term
[Fig.~\ref{fig:dia1}(a)]; (2) $t$($u)$-channel vector meson exchange
[Fig.~\ref{fig:dia1}(b)]; (3) $s$-channel vector meson exchange
[Fig.~\ref{fig:dia1}(c)]; (4) box diagram with intermediate
pseudoscalars [Fig.~\ref{fig:dia1}(d)]. 
The corresponding diagram to the one in 
Fig.~\ref{fig:dia1}(d) with crossed pions for $\rho\rho$ scattering
was shown in Ref.~\cite{Molina:2008jw} to provide
much smaller contribution than the direct box diagram [Fig. 1(d)] and, hence, we
ignore it here. Similarly in Ref.~\cite{Molina:2008jw} the contribution
of box diagrams with intermediate vector mesons involving anomalous
couplings was also found to be small and we shall omit them
in the present work as well.

In our approach, the first two diagrams play the most
important role in the formation of  resonances. The $s$-channel
vector meson exchange is mostly of $p$-wave nature. In the case of
the strangeness=1 channel, an $s$-wave contribution appears, which is
proportional to the differences between the initial (final) vector
meson masses and is found to be numerically negligible compared to
the sum of the contact mechanism and the $t$($u$)-channel vector meson exchange
mechanism. The box diagram depends somewhat on a form factor that
we shall discuss later on. The real part of the amplitude is small compared to
the sum of the four-vector-contact amplitude and the $t$($u$) channel
vector-exchange amplitude, but the imaginary part
is relatively large because there is a large phase space for the decay into two
pseudoscalars, as has been explicitly shown in Ref.~ \cite{Molina:2008jw},
where
cancellations of the real part with that from the box diagram involving anomalous couplings was also found.
Thus, we keep only its imaginary part.

We adopt the hidden-gauge formalism, consistent with chiral
symmetry, to describe the interactions between the vector mesons and
those between the vectors and the
pseudoscalars~\cite{Bando:1984ej,Bando:1987br}. The hidden-gauge
Lagrangian is
\begin{equation}\label{eq:laghg}
\mathcal{L}=-\frac{1}{4}\langle \bar{V}_{\mu\nu}\bar{V}^{\mu\nu}\rangle
+\frac{1}{2}M_v^2 \langle [V_\mu-(i/g)\Gamma_\mu]^2\rangle,
\end{equation}
where
\begin{equation*}
\bar{V}_{\mu\nu}=\partial_\mu V_\nu-\partial_\nu V_\mu-ig[V_\mu,V_\nu],
\end{equation*}
\begin{equation*}
\Gamma_\mu=\frac{1}{2}\left\{
u^\dagger [\partial_\mu -i(v_\mu+a_\mu)]u +
u[\partial_\mu-i(v_\mu-a_\mu)]u^\dagger\right\},
\end{equation*}
and $\langle\rangle$ stands for the trace in the SU(3) flavor space.
$V_\mu$ represents the vector nonet:
 \begin{equation}
\renewcommand{\tabcolsep}{1cm}
\renewcommand{\arraystretch}{2}
V_\mu=\left(
\begin{array}{ccc}
\frac{\omega+\rho^0}{\sqrt{2}} & \rho^+ & K^{*+}\\
\rho^- &\frac{\omega-\rho^0}{\sqrt{2}} & K^{*0}\\
K^{*-} & \bar{K}^{*0} &\phi
\end{array}
\right)_\mu,
\end{equation}
while $u^2=U=\exp\left(\frac{i\sqrt{2}\Phi}{f}\right)$ with
$\Phi$ the octet of the pseudoscalars
\begin{equation}
\renewcommand{\tabcolsep}{1cm}
\renewcommand{\arraystretch}{2}
\Phi=\left(
\begin{array}{ccc}
\frac{\eta}{\sqrt{6}}+\frac{\pi^0}{\sqrt{2}} & \pi^+ & K^+\\
\pi^- &\frac{\eta}{\sqrt{6}}-\frac{\pi^0}{\sqrt{2}} & K^{0}\\
K^{-} & \bar{K}^{0} &-\sqrt{\frac{2}{3}}\eta
\end{array}
\right).
\end{equation}
The value of the coupling constant $g$ of the Lagrangian [Eq.~(\ref{eq:laghg})]
is
\begin{equation}
g=\frac{M_V}{2f},
\end{equation}
with $M_V$ the vector meson mass and $f=93$ MeV the pion decay constant.

The Lagrangian of Eq.~(\ref{eq:laghg}) provides the following two
interactions:
\begin{equation}
\mathcal{L}_\mathrm{VVVV}=\frac{1}{2}g^2\langle [V_\mu,V_\nu]V^\mu V^\nu\rangle,
\end{equation}
\begin{eqnarray}\label{eq:3v}
\mathcal{L}_{VVV}&=&ig\langle (\partial_\mu V_\nu-\partial_\nu
V_\mu)V^\mu V^\nu\rangle\nonumber\\
&=&ig\langle V^\mu \partial_\nu V_\mu V^\nu-\partial_\nu V_\mu V^\mu
V^\nu \rangle\nonumber\\
&=&ig\langle (V^\mu \partial_\nu V_\mu -\partial_\nu V_\mu V^\mu)
V^\nu)\rangle.
\end{eqnarray}
The first one is responsible for the four-vector-contact interaction and the second one
leads to the $s$, $t$, $u$-channel vector-exchange mechanisms.

To calculate the box diagram, one needs the vector-pseudoscalar-pseudoscalar interaction,
which is also provided by Eq.~(\ref{eq:laghg}) as
\begin{equation}\label{eq:vpp}
\mathcal{L}_{V\Phi\Phi}=-ig\langle
V_\mu[\Phi,\partial^\mu\Phi]\rangle.
\end{equation}

With the above vertices, one can then calculate the tree-level transition amplitudes for
 each strangeness and isospin channel.
 With the interaction of two spin one particles, the final state could have
either spin 0, spin 1, or spin 2. One then has the following strangeness, isospin, and spin channels:
(0,0,0), (0,0,1), (0,0,2), (0,1,0), (0,1,1), (0,1,2), (0,2,0), (0,2,1), (0,2,2),
(1,1/2,0), (1,1/2,1), (1,1/2,2), (1,3/2,0), (1,3/2,1), (1,3/2,2), (2,0,0), (2,0,1), (2,0,2),
(2,1,0), (2,1,1), and (2,1,2). In total, there are 21 channels. Proceeding further,
we will see that not in all of these channels the vector meson-vector meson interaction leads to
 resonances.

An important ingredient in the Bethe-Salpeter equation method is the
on-shell evaluation of the transition amplitudes $V$, which reduces
the coupled channel integral equations to coupled channel algebraic
equations. This can be justified using various methods, such as
through a dispersion relation on $T^{-1}$ after imposing unitarity
~\cite{Oller:1998zr,Oller:2000fj}, or in a more transparent way, writing
$V(q^2)\simeq V(m^2)+\frac{\partial V}{\partial q^2} (q^2-m^2)$. The
off-shell part of the amplitude then cancels one vector meson
propagator, leading to a tadpole kind of diagram. This diagram gets
canceled with genuine tadpole diagrams from the same chiral
Lagrangian, or, can be taken into account by redefining the
couplings of the original transition amplitude. In any case, one can
evaluate the transition amplitudes on shell.

Since we are only interested in the energy region close to the
vector meson-vector meson threshold, one can safely ignore the
three-momenta of the external vector mesons relative to their
masses and, hence, the zero component of their polarization
vectors. With the above mentioned on-shell factorization, as
explained in detail in Ref.~\cite{Molina:2008jw}, one can prove that,
after neglecting corrections of the order $|\vec{q}|^2/M_V^2$, the
vector meson propagators in the loops of the Bethe-Salpeter series
can be simplified as
\begin{equation}
\frac{\delta_{ij}}{q^2-M^2_V+i\epsilon}
\end{equation}
with $i$, $j$ the spacial indices of the polarization vectors. On
the other hand, the propagator for the vector mesons exchanged in
the $t$ and $u$ channels entering the evaluation of the tree-level
transition amplitudes is given by
\begin{equation}
-g^{\mu\nu}\frac{1}{q^2-M_V^2+i\epsilon}.
\end{equation}

With the approximations mentioned above of
neglecting the three-momenta of the vector mesons versus their masses, 
the projection operators into spin 0, 1, and 2, in terms of
the four polarization vectors, are~\cite{Molina:2008jw}
\begin{eqnarray}\label{eq:projector}
\mathcal{P}^{(0)}&=&\frac{1}{3}\epsilon(1)\cdot\epsilon(2)\,\epsilon(3)\cdot \epsilon(4),\nonumber\\
\mathcal{P}^{(1)}&=&\frac{1}{2}\big[\epsilon(1)\cdot\epsilon(3)\,\epsilon(2)\cdot\epsilon(4)-
\epsilon(1)\cdot\epsilon(4)\,\epsilon(2)\cdot\epsilon(3)\big],\nonumber\\
\mathcal{P}^{(2)}&=&\frac{1}{2}\big[\epsilon(1)\cdot\epsilon(3)\,\epsilon(2)\cdot\epsilon(4)+
\epsilon(1)\cdot\epsilon(4)\,\epsilon(2)\cdot\epsilon(3)\big]\nonumber\\
&&-\frac{1}{3}\epsilon(1)\cdot\epsilon(2)\,
\epsilon(3)\cdot\epsilon(4).
\end{eqnarray}

In the following, we explain how to calculate the
three-kinds of tree-level transition amplitudes, i.e.,
the four-vector-contact amplitude [Fig.~\ref{fig:dia1}(a)], the $t(u)$-channel vector-exchange 
amplitude [Fig.~\ref{fig:dia1}(b)], and the box amplitude [Fig.~\ref{fig:dia1}(d)].

\subsection{Four-vector-contact term}
With the spin projectors and the Lagrangian $\mathcal{L}_{VVVV}$,
one can easily obtain the $V_{ij}$'s for different strangeness,
isospin, and spin channels. The results are summarized in Tables
\ref{table:vvvv1}-\ref{table:vvvv13} in Appendix A. One thing to
note is that for each pair of identical particles a factor of
$\frac{1}{\sqrt{2}}$ has to be multiplied, i.e., the unitarity
normalization, which originates from the fact that
\begin{equation}
\frac{1}{2}\sum_q |I(\vec{q})I(-\vec{q})\rangle\langle I(\vec{q})I(-\vec{q})|=1,
 \end{equation}
 where $I$ denotes the identical particle~\cite{Oller:1997ti}. One has
 to keep in mind that the unitarity
 normalization has to be used to calculate the
 $t$($u$)-channel vector-exchange diagrams and the box diagrams as well.

To obtain the amplitudes in isospin space, we use the following phase convention:
\begin{equation}
\rho^+=-|1,+1\rangle,\quad
K^{*-}=-|1/2,-1/2\rangle.
\end{equation}
\subsection{Vector exchange in $t$($u)$ channel}
To calculate the $t$($u$)-channel vector meson exchange diagrams,
one has to project the vertices into $s$ wave. This can be done
by the following replacements:
\begin{eqnarray}
k_1\cdot k_2&=&\frac{s-M_1^2-M_2^2}{2},\nonumber\\
k_1\cdot k_3&=&k_1^0k_3^0-\vec{p}\cdot\vec{q}\rightarrow\frac{(s+M_1^2-M_2^2)(s+M_3^2-M_4^2)}{4s},\nonumber\\
k_1\cdot k_4&=&k_1^0k_4^0+\vec{p}\cdot\vec{q}\rightarrow\frac{(s+M_1^2-M_2^2)(s-M_3^2+M_4^2)}{4s},\nonumber\\
k_2\cdot k_3&=&k_2^0k_3^0+\vec{p}\cdot\vec{q}\rightarrow\frac{(s-M_1^2+M_2^2)(s+M_3^2-M_4^2)}{4s},\nonumber\\
k_2\cdot k_4&=&k_2^0k_4^0-\vec{p}\cdot\vec{q}\rightarrow\frac{(s-M_1^2+M_2^2)(s-M_3^2+M_4^2)}{4s},\nonumber\\
k_3\cdot k_4&=&\frac{s-M_3^2-M_4^2}{2},\nonumber
 \end{eqnarray}
 where $\rightarrow$ means the projection over $s$ wave,
 and  $k_1=(k^0_1,\vec{p})$, $k_2=(k^0_2,-\vec{p})$,
 $k_3=(k^0_3,\vec{q})$, $k_4=(k^0_4,-\vec{q})$ are
 the four-momenta of the particles 1, 2, 3, and 4 with
 masses $M_1$, $M_2$, $M_3$, and $M_4$. 

The last expression of
 Eq.~(\ref{eq:3v}) is particularly suitable for the calculation of the
 vertices. Indeed, the vector field $V^\nu$ must correspond
 necessarily to the exchanged vector meson. If it were an
 external vector meson, the $\nu$ must be spatial as we mentioned
 and then $\partial_\nu$ leads to a three-momentum of an external
 vector, which is neglected in the present approach. Given the
 structure of the last expression in Eq.~(\ref{eq:3v}) one can
 easily see that all terms corresponding to the $t$ channel
 ($1+2\rightarrow3+4$) have the type
 \begin{equation}
 (k_1+k_3)\cdot (k_2+k_4)\;\epsilon_1\cdot\epsilon_3
 \epsilon_2\cdot\epsilon_4,
 \end{equation}
 while those corresponding to $u$-channel diagrams
 $(1+2\rightarrow4+3)$ have the structure
 \begin{equation}
 (k_1+k_4)\cdot(k_2+k_3)\;\epsilon_1\cdot\epsilon_4
 \epsilon_2\cdot\epsilon_3.
 \end{equation}
It is interesting to note that the above structures of the $t(u)$ channel
vector-exchange contributions, together with
the structures of the projection operators [Eq.~(\ref{eq:projector})], imply that they contribute equally to spin=0 and
spin=2 states.

 The resulting tree-level transition amplitudes are summarized
 in Tables \ref{table:3v1}-\ref{table:3v10} in Appendix A.

 \subsection{Box diagrams}
 The box diagrams provide a mechanism for the dynamically generated resonances
 to decay into two pseudoscalars. With the $\mathcal{L}_{V\Phi\Phi}$
 Lagrangian of Eq.~(\ref{eq:vpp}) and our assumption that the external particles have
 small three-momenta, these diagrams can be easily calculated, as shown in Ref.~\cite{Molina:2008jw} 
and explained in the following.

\begin{figure}[t]
\includegraphics[scale=0.6]{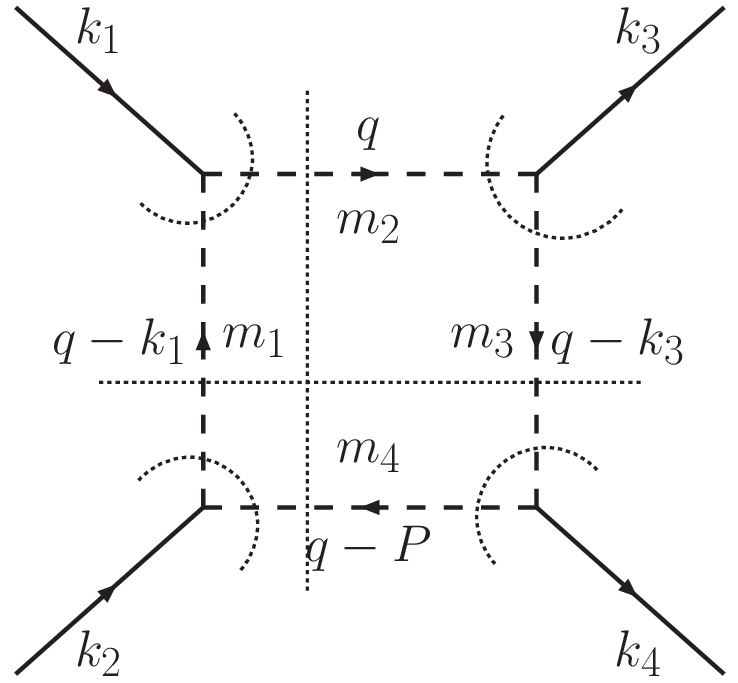}
\caption{Kinematics of a generic box diagram, with
$m_1$, $m_2$, $m_3$, and $m_4$ denoting the masses of intermediate pseudoscalars and
$k_1$, $k_2$, $k_3$, and $k_4$ the four-momenta of the vector particles.\label{fig:diabox}}
\end{figure}

\begin{figure}[t]
\includegraphics[scale=0.4]{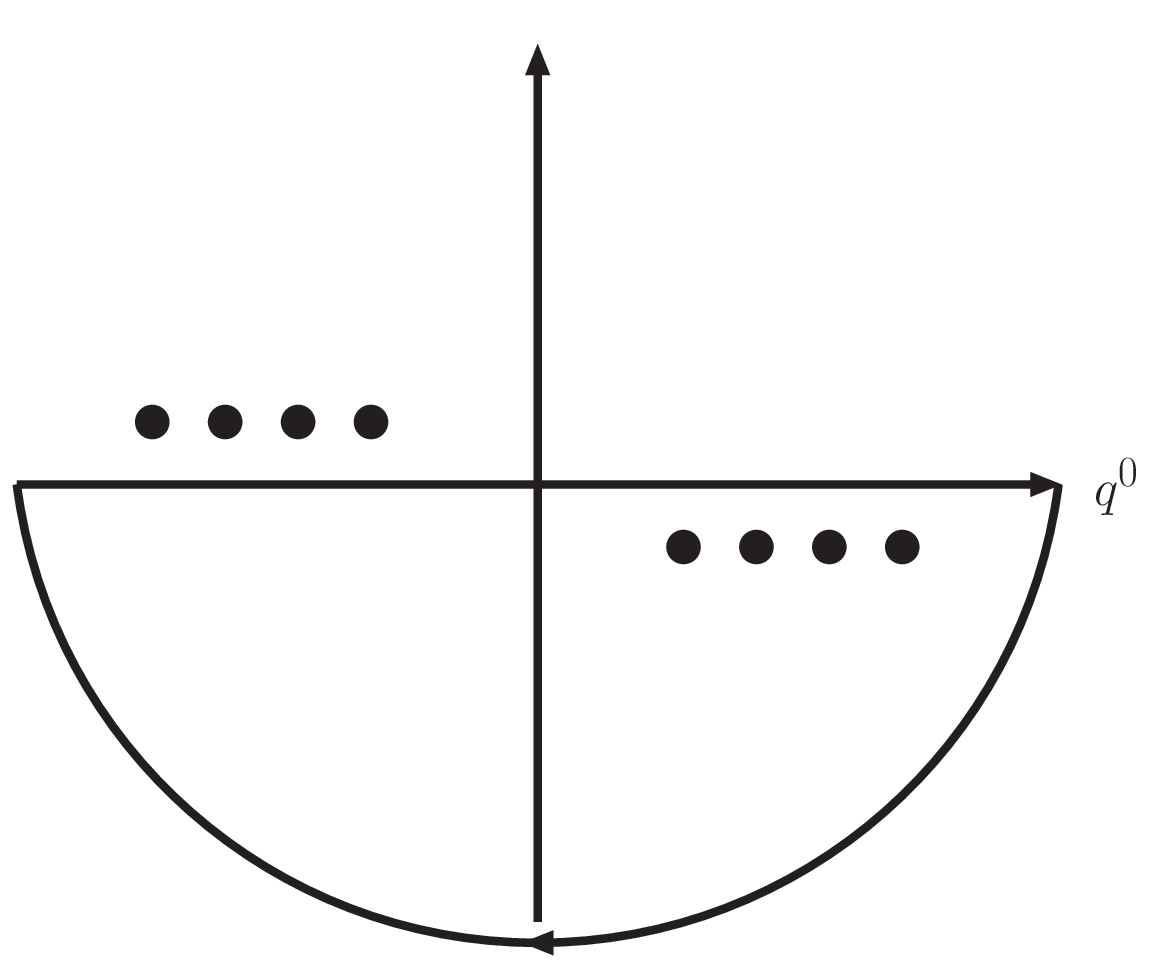}
\caption{The contour to evaluate a generic box diagram.\label{fig:contour}}
\end{figure}

 The box diagrams have the following generic structure (with the notations shown
in Fig.~\ref{fig:diabox})
 \begin{eqnarray}
 V_b&\sim&C \int\frac{d^4q}{(2\pi)^4} \epsilon_1\cdot(2q-k_1) \epsilon_2\cdot (2q-k_3)\\
 && \times \epsilon_3\cdot(2q-k_3-P)\epsilon_4\cdot(2q-k_1-P)\nonumber\\
 &&\times\frac{1}{(q-k_1)^2-m_1^2+i\epsilon}
 \frac{1}{q^2-m_2^2+i\epsilon}\nonumber\\
 &&\times\frac{1}{(q-k_3)^2-m_3^2+i\epsilon}\frac{1}{(q-P)^2-m_4^2+i\epsilon},\nonumber
 \end{eqnarray}
 where $C$ is the coupling of a certain transition.
 With the approximation of neglecting the three-momenta of the external particles, this can be simplified as

  \begin{eqnarray}
 V_b&\sim&C' \int\frac{d^4q}{(2\pi)^4} \epsilon_1^i\epsilon_2^j\epsilon_3^m\epsilon_4^n q^i q^j q^m q^n\nonumber\\
 &&\times\frac{1}{(q-k_1^0)^2-m_1^2+i\epsilon}
 \frac{1}{q^2-m_2^2+i\epsilon}\nonumber\\
 &&\times\frac{1}{(q-k_3^0)^2-m_3^2+i\epsilon}\frac{1}{(q-P^0)^2-m_4^2+i\epsilon}\nonumber\\
 &=&C' G,
 \end{eqnarray}
 with $C'=16C$. To calculate this integral, we first integrate the $q^0$ variable by use
 of the residue theorem and close the integral below, as shown in Fig.~\ref{fig:contour},
 \begin{equation}
 G=(-2\pi i)\frac{1}{2\pi}\int\frac{d^3q}{(2\pi)^3}\epsilon_1^i\epsilon_2^j\epsilon_3^m\epsilon_4^n q^i q^j q^m q^n\times \frac{G_\mathrm{n}}{G_\mathrm{d}}
 \end{equation}
 with
 \begin{eqnarray}\label{eq:Gd}
 G_\mathrm{d}&=&\frac{1}{2 \omega _1 \omega _2 \omega _3\omega _4}
 \frac{1}{\left(-P^0 -\omega _2-\omega _4\right)}\nonumber\\
 &\times&\frac{1}{\left(k_1^0+\omega _1+\omega _2\right)}
 \frac{1}{\left(k_3^0+\omega _2+\omega _3\right)}\nonumber\\
 &\times&
 \frac{1}{\left( k_4^0+\omega _3+\omega _4\right)}
 \frac{1}{\left(k_2^0+\omega _1+\omega
   _4\right)}\nonumber\\
   &\times& \frac{1}{\left(k_1^0-\omega _1-\omega _2 +i \epsilon\right)}
   \frac{1}{\left(k_3^0-\omega _2-\omega _3+i \epsilon\right)}\nonumber\\
   &\times&\frac{1}{\left(k_2^0-\omega _1-\omega _4+i \epsilon\right)}
   \frac{1}{\left(k_4^0-\omega _3-\omega _4+i \epsilon\right)}\nonumber\\
   &\times& \frac{1}{\left( P^0-\omega _2-\omega _4+i \epsilon\right)}\nonumber\\
   &\times& \frac{1}{\left(k_1^0-k_3^0-\omega _1-\omega _3+i \epsilon\right)}
   \frac{1}{\left(k_3^0+k_1^0-\omega _1-\omega _3+i \epsilon\right)}\nonumber,
 \end{eqnarray}
 where different cuts contributing to the imaginary part of the integral can be clearly seen 
(see also the dotted lines in Fig.~\ref{fig:diabox}), and
 $\omega_1=\sqrt{q^2+m_1^2}$, $\omega_2=\sqrt{q^2+m_2^2}$,
 $\omega_3=\sqrt{q^2+m_3^2}$, $\omega_4=\sqrt{q^2+m_4^2}$,
 $k_1^0=\frac{s+M_1^2-M_2^2}{2\sqrt{s}}$, $k_2^0=\frac{s+M_2^2-M_1^2}{2\sqrt{s}}$,
 $k_3^0=\frac{s+M_3^2-M_4^2}{2\sqrt{s}}$, $k_4^0=\frac{s+M_4^2-M_3^2}{2\sqrt{s}}$, and $P^0=\sqrt{s}$,
 where $m_1$, $m_2$, $m_3$, and $m_4$ are the masses of intermediate pseudoscalars, $M_1$, $M_2$, $M_3$, and
 $M_4$ are the masses of the initial and final vector mesons,
and $\sqrt{s}$ is the center of mass of energy
 of the vector-vector pair. $G_\mathrm{n}$ is also a function of
 these variables, whose explicit form is given in Appendix C.

 Since
 \begin{eqnarray}
\int d^3q\; q_i q_j q_m q_n f(q)&=& \frac{1}{15}\int d^3q\; q^4 f(q)\\
 &&\hspace{0cm}\times(\delta_{ij}\delta_{mn}+\delta_{im}\delta_{jn}+\delta_{in}\delta_{jm})\nonumber,
 \end{eqnarray}
 the four-point integral $G$ becomes
 \begin{eqnarray}
 G&=&(-i)\frac{1}{15}\frac{1}{2\pi^2}\int dq\;q^6\frac{G_\mathrm{n}}{G_\mathrm{d}}\times
 \Big[\epsilon(1)\cdot\epsilon(2)\epsilon(3)\cdot\epsilon(4)\nonumber\\
 &&\hspace{0.5cm}
 +\epsilon(1)\cdot\epsilon(3)\epsilon(2)\cdot\epsilon(4) +\epsilon(1)\cdot\epsilon(4)\epsilon(2)\cdot\epsilon(3)\Big]\nonumber\\
 &=&(-i)\frac{1}{15}\frac{1}{2\pi^2}\int dq\;q^6\frac{G_\mathrm{n}}{G_\mathrm{d}}\times
 (5 P^{(0)}+2P^{(2)}).
 \end{eqnarray}
As one can see from the above result, there is no contribution to spin=1 channels from 
the box diagrams. This should be the case since two vectors in $L=0$ have positive parity.
To have $J=1$ with two pseudoscalars one needs $L'=1$ in the two pseudoscalars system,
which, however, has negative parity.
 It is interesting to note that the box diagrams contribute 2.5 times more to
 the spin zero states than to the spin 2 states. This is one of the
 reasons why the scalar resonances develop a larger width than the
 tensor ones. The fact that the tensor resonances are more bound
 than the scalar ones reinforces this trend.


The explicit forms of the transition amplitudes are given in 
Appendix B with the following structure:
\begin{eqnarray}\label{eq:vbox}
v_{i,j}&=&\sum c G_4(m_1,m_2,m_3,m_4,s,k_1^0,k_2^0,k_3^0,k_4^0),\nonumber\\
   &+&\sum \tilde{c} G_4(m_1,m_2,m_3,m_4,s,k_1^0,k_2^0,k_4^0,k_3^0),
\end{eqnarray}
with $c$ and $\tilde{c}$ the couplings and $G_4$ the
four-point function defined as
\begin{equation}
G_4=\frac{1}{15}\frac{1}{2\pi^2}\int dq\,q^6\frac{G_n}{G_d}
\end{equation}
with $G_d$ given in Eq.~(\ref{eq:Gd}) and $G_n$ given in 
Appendix C.

As in Ref.~\cite{Molina:2008jw}, we evaluate the $G_4$ loop function with
a cutoff of $\Lambda\sim1$ GeV. To avoid the appearance of double poles, we
replace the $k_i^0$'s in the denominator $G_d$ by
\begin{equation}
k^0_1\rightarrow k^0_1+\frac{\Gamma_1}{4},\quad
k^0_3\rightarrow k^0_3-\frac{\Gamma_3}{4},
\end{equation}
\begin{equation}
k^0_2\rightarrow k^0_2+\frac{\Gamma_2}{4},\quad
k^0_4\rightarrow k^0_4-\frac{\Gamma_4}{4}.
\end{equation}
This was found to be a good approximation in  Ref.~\cite{Molina:2008jw}
to the more accurate method of removing the double poles, which
consists in making a convolution over the mass distributions of the
external vector mesons to account for their widths.

We also multiply the vertices by the following form factors:
\begin{equation}\label{eq:form1}
F_1(q^2)=\frac{\Lambda_b^2-m_1^2}{\Lambda_b^2-(k_1^0-q^0)^2+|\vec{q}|^2},
\end{equation}
\begin{equation}\label{eq:form2}
F_3(q^2)=\frac{\Lambda_b^2-m_3^2}{\Lambda_b^2-(k_3^0-q^0)^2+|\vec{q}|^2},
\end{equation}
with $q^0=\frac{s+m_2^2-m_4^2}{2\sqrt{s}}$, $\vec{q}$ the running variable,
and $\Lambda_b=1.4$ GeV~\cite{Molina:2008jw}. These form factors
are inspired by the fact that the largest piece of the imaginary
part of $G_4$ comes from the cut at $P^0=\omega_2+\omega_4$ and inspired
by the empirical form factors used in the decay of vector mesons~\cite{Titov:2000bn,Titov:2001yw}.
The final form of the four-point function is then
\begin{equation}\label{eq:4ploop}
G_4=\frac{1}{15}\frac{1}{2\pi^2}\int_0^\Lambda dq\,q^6  \frac{G_n}{G_d}F_1(q^2)^2 F_3(q^2)^2.
\end{equation}

Using the explicit expressions of $v_{i,j}$ of Eq.~(\ref{eq:vbox}) in Appendix B, we can also calculate
the partial decay widths into two pseudoscalars selecting only the particular channels. Technically,
this implies keeping only the relevant terms in each $v_{i,j}$ of Appendix B. For
instance, take the case of the strangeness=0, isospin=0, and spin=0 channel as one example.
If we want to have the $\pi\pi$ decay mode we keep only the terms that have $m_\pi$, $m_\pi$ in
the second and fourth arguments of each $\tilde{G}$ ($\tilde{G}(u)$) function.

\section{results and discussions}
Since the vector mesons, particularly the $\rho$ and the $K^*$, are rather
broad, one has to take into account their widths. We follow Ref.~\cite{Molina:2008jw}
and convolute the vector-vector $G$ function with the mass distributions of
the two vector mesons, i.e., by replacing the $G$ function appearing in
the Bethe-Salpeter equation [Eq.~(\ref{eq:Gsharp})] by $\tilde{G}$
\begin{eqnarray}
\tilde{G}(s)&=&\frac{1}{N^2}\int\limits^{(M_1+2\Gamma_1)^2}\limits_{(M_1-2\Gamma_1)^2}
d\tilde{m}^2_1\left(-\frac{1}{\pi}\right)\mathrm{Im}\frac{1}{\tilde{m}^2_1-M_1^2+i\tilde{\Gamma}_1\tilde{m}_1}\nonumber\\
&&\times\int\limits^{(M_2+2\Gamma_2)^2}\limits_{(M_2-2\Gamma_2)^2}
d\tilde{m}^2_2\left(-\frac{1}{\pi}\right)\mathrm{Im}\frac{1}{\tilde{m}^2_2-M_2^2+i\tilde{\Gamma}_2\tilde{m}_2}
\nonumber\\
&&\times G(s,\tilde{m}_1^2,\tilde{m}_2^2)
\end{eqnarray}
with
\begin{eqnarray}
N^2&=&\int\limits^{(M_1+2\Gamma_1)^2}\limits_{(M_1-2\Gamma_1)^2}
d\tilde{m}^2_1\left(-\frac{1}{\pi}\right)\mathrm{Im}\frac{1}{\tilde{m}^2_1-M_1^2+i\tilde{\Gamma}_1\tilde{m}_1}\nonumber\\
&&\times\int\limits^{(M_2+2\Gamma_2)^2}\limits_{(M_2-2\Gamma_2)^2}
d\tilde{m}^2_2\left(-\frac{1}{\pi}\right)\mathrm{Im}\frac{1}{\tilde{m}^2_2-M_2^2+i\tilde{\Gamma}_2\tilde{m}_2},\nonumber
\end{eqnarray}
where $M_1$, $M_2$, $\Gamma_1$, and $\Gamma_2$ are the masses and
widths of the two vector mesons in the loop. We only take into
account the widths of the $\rho$ and the $K^*$. In the case of the
$\omega$ or $\phi$, one or both of the kernels of these integrals
will reduce to a delta function $\delta(\tilde{m}^2-M^2)$. The
$\tilde{\Gamma}_i$ function is energy dependent and has the form of
\begin{equation}
\tilde{\Gamma}(\tilde{m})=\Gamma_0\frac{q^3_\mathrm{off}}{q^3_\mathrm{on}}\Theta(\tilde{m}-m_1-m_2)
\end{equation}
with
\begin{equation}
q_\mathrm{off}=\frac{\lambda(\tilde{m}^2,m_\pi^2,m_\pi^2)}{2\tilde{m}},\quad
q_\mathrm{on}=\frac{\lambda(M_\rho^2,m_\pi^2,m_\pi^2)}{2 M_\rho}
\end{equation}
and $m_1=m_2=m_\pi$ for the $\rho$ or
\begin{equation}
q_\mathrm{off}=\frac{\lambda(\tilde{m}^2,m_K^2,m_\pi^2)}{2\tilde{m}},\quad
q_\mathrm{on}=\frac{\lambda(M_{K^*}^2,m_K^2,m_\pi^2)}{2 M_{K^*}},
\end{equation}
$m_1=m_\pi$ and $m_2=m_K$ for the $K^*$, where $\lambda$ is the
K\"allen function, $\lambda(x,y,z)=(x-y-z)^2-4yz$, and $\Gamma_0$ is the 
nominal width of the $\rho$ or the $K^*$.

To regularize the loop functions, one can use either the cutoff
method with a natural cutoff of $\sim1$ GeV or the dimensional regularization
method with $a\sim-2$ for meson-baryon scattering~\cite{Oller:2000fj}.
This means that by using these parameter values one should get the basic physics,
providing a global description of the resonances generated dynamically in the approach.
This is indeed the case here. Yet, in order to take into account possible correcting
terms in the approach, we perform a fine-tuning of these parameters, such as
to get a few resonances more precisely. Then, the results for other resonances are
predictions. In practice, we adopt the following three-steps approach:
\begin{enumerate}
\item First we use the cutoff method with
$\Lambda\sim1$ GeV to obtain the amplitudes on the real axis.

 \item Once peaks
and bumps are observed, and persist with reasonable adjustments of the value of the cutoff $\Lambda$,
we then use the dimensional regularization method
with $\mu=1000$ MeV and $a$ adjusted to reproduce the cutoff results.
More specifically, we reproduce the real part of the rho-rho loop function
at the two $\rho$ threshold. This gives $a=-1.65$.

\item Then we fine-tune the $a$'s for different isospin channels to
fix the masses of some well-known resonances. In the present work,
we use the masses of the $f_2(1270)$, the $f'_2(1525)$, and the
$K^*_2(1430)$ for this purpose. This leads to $a_{\rho\rho}=-1.636$,
$a_{K^*\bar{K}^*}=-1.726$, $a_{\rho K^*}=-1.85$. For the rest of the
channels involving $\omega$ or $\phi$, in the strangeness=0 channel we use
$a_i=a_{\rho\rho}=-1.65$; in the strangeness=1 channel we use
$a_i=a_{\rho K^*}=-1.85$; and in the strangeness=2 channel,
we use $a_i=a_{K^*\bar{K}^*}=-1.726$. These channels play a secondary role and moderate
changes of these parameters barely affect the results. Hence, in practice, we are
fine-tuning three parameters.

 We should mention that
our main conclusions would remain the same if we had  used, for instance,
the same value of 
$a_i=-1.85$ for all the channels, and we find only moderate changes in
the masses of the resonances. For instance, with this choice of $a_i$,
we would obtain the $f_2(1270)$ at $(1206,-i0)$ MeV on the complex plane without
including the box diagrams, compared to
$(1275,-i1)$ MeV with the fine-tuned subtraction constants, and the 
$1^-(0^{++})$ state at $(1770,-i50)$ MeV instead of 
$(1780,-i66)$ MeV (see Tables \ref{table:res1} and \ref{table:res2}). 
This means that we get the bulk of the resonances using a natural substraction constant (cutoff) for the effective 
field theory. Once this is done, fine-tuning of parameters 
will provide a better description of these
resonances. Since we get 11 dynamically generated resonances and
have fun-tunned three parameters to get the masses of the 
three resonances, we are making predictions for
eight of them.

As to the total width of the resonances, they are sensitive to
the form factors given in Eqs.~(\ref{eq:form1},\ref{eq:form2}). The form factors used were inspired by
the study of Refs.~\cite{Titov:2000bn,Titov:2001yw} and the precise value for $\Lambda_b$
was taken from the study of Ref.~\cite{Molina:2008jw}. Later in this section we mention
the sensitivity of the width to changes in the $\Lambda_b$ value.
Once again we can invoke the same fine-tuning strategy discussed above and say that a certain
value of $\Lambda_b$ is taken to get the total width of one of the fitted resonances, such that the widths of the others are predictions.

We should also note that the couplings of the resonances to the coupled channels are rather
independent of the $\Lambda_b$ parameter, which was already found in Ref.~\cite{Nagahiro:2008um}.

Finally, let us mention that our approach also predicts branching ratios to different channels.
The parameters of the theory have not been fine-tunned to these observables and, hence, all the
branching ratios obtained are genuine predictions of our approach, which seem to be consistent with
data as shown in the following sections.
\end{enumerate}

The combination of the cutoff method and the dimensional regularization method has
the following advantage: The use of the cutoff method is physically more transparent: the value of
the cutoff should be around 1 GeV in order for the results to make sense. The use of
the dimensional regularization method, on the other hand, enables one to go to the second Riemann sheet to
obtain the pole positions and the residues. The results shown below are obtained in the
dimensional regularization scheme. For the masses and widths of the vector mesons, we use
the following values~\cite{Amsler:2008zz}: $M_\rho=775.49$ MeV, $\Gamma_\rho=149.4$ MeV,
$M_{K^*}=893.83$ MeV, $\Gamma_{K^*}=50.55$ MeV, $M_\omega=782.65$ MeV,
$M_\phi=1019.455$ MeV. For the masses of the pseudoscalars,
the following values are used: $m_\pi=138.04$ MeV,
$m_K=495.66$ MeV,  $m_\eta=547.51$ MeV~\cite{Amsler:2008zz}. The coupling constant $g=\frac{M_V}{2f}$ is
evaluated with $M_V=M_\rho$ and $f=93$ MeV. Of course, one could also use
an averaged mass for $M_V$ and an averaged $f$. In this case, both the numerator
and the denominator will become somewhat larger, and the ratio is only slightly changed.
Otherwise, in the potentials and in the $G(s)$ functions we have used the
physical masses of the particles, as mentioned above. This, in particular, the large $\phi$
and $\rho$ mass difference, introduces a certain source of SU(3) breaking which
might not be the only one present in the problem. However, the consideration of the physical masses is absolutely
necessary to guarantee unitarity in coupled channels and to respect the positions of the thresholds, and this is the main reason to stick to physical masses in our approach.

The free parameters are then the subtraction constants used to
regularize the vector-vector loop functions. In fact, the values can
be different for each isospin channel, and may even be different for
different spins, but only slight changes can be
expected~\cite{Molina:2008jw}. Since the main purpose of this paper
is to extend the work of Ref.~\cite{Molina:2008jw} and to see whether in
other strangeness-isospin-spin channels resonances can be
dynamically generated, we do not use that freedom to fine-tune all
the subtraction constants, which only leads to small changes in the
masses of the resonances obtained.
\begin{table*}[htpb]
      \renewcommand{\arraystretch}{1.0}
     \setlength{\tabcolsep}{0.1cm}
     \centering
     \caption{Pole positions and residues in the strangeness=0 and isospin=0 channel.
             All quantities are in units of MeV.\label{table:res1}}
     \begin{tabular}{c|ccccc}
     \hline\hline
      \multicolumn{6}{c}{$(1512,-i26)$ [spin=0]}\\\hline
       &  $K^*\bar{K}^*$ & $\rho\rho$ & $\omega\omega$  & $\omega\phi$  & $\phi\phi$ \\\hline
 $g$ &  $(1208,-i419)$ & $(7920,-i1071)$ & $(-39,i31)$ & $(33,-i43)$ & $(12,i24)$ \\
   \hline\hline
      \multicolumn{6}{c}{$(1726,-i14)$ [spin=0]}\\\hline
       &  $K^*\bar{K}^*$ & $\rho\rho$ & $\omega\omega$  & $\omega\phi$  & $\phi\phi$ \\\hline
 $g$ &  $(7124,i96)$ & $(-1030,i1086)$ & $(-1763,i108)$ & $(3010,-i210)$ & $(-2493,-i204)$ \\
    \hline\hline
      \multicolumn{6}{c}{$(1802,-i39)$ [spin=1]}\\\hline
       &  $K^*\bar{K}^*$ & $\rho\rho$ & $\omega\omega$  & $\omega\phi$  & $\phi\phi$ \\\hline
 $g$ &  $(8034,-i2542)$ & $0$ & $0$ & $0$ & $0$ \\
    \hline\hline
      \multicolumn{6}{c}{$(1275,-i1)$ [spin=2]}\\\hline
       &  $K^*\bar{K}^*$ & $\rho\rho$ & $\omega\omega$  & $\omega\phi$  & $\phi\phi$ \\\hline
 $g$ &  $(4733,-i53)$ & $(10889,-i99)$ & $(-440,i7)$ & $(777,-i13)$ & $(-675,i11)$ \\
     \hline\hline
      \multicolumn{6}{c}{$(1525,-i3)$ [spin=2]}\\\hline
       &  $K^*\bar{K}^*$ & $\rho\rho$ & $\omega\omega$  & $\omega\phi$  & $\phi\phi$ \\\hline
 $g$ &  $(10121,i101)$ & $(-2443,i649)$ & $(-2709,i8)$ & $(5016,-i17)$ & $(-4615,i17)$ \\\hline\hline
    \end{tabular} 
 \end{table*}

  \begin{table*}[htpb]
      \renewcommand{\arraystretch}{1.5}
     \setlength{\tabcolsep}{0.1cm}
     \centering
     \caption{The same as Table \ref{table:res1}, but for the strangeness=0 and isospin=1 channel.\label{table:res2}}
     \begin{tabular}{c|cccc}
     \hline\hline
      \multicolumn{5}{c}{$(1780,-i66)$ [spin=0]}\\\hline
       &  $K^*\bar{K}^*$ & $\rho\rho$ & $\rho\omega$  & $\rho\phi$ \\\hline
 $g$ &  $(7525,-i1529)$ & $0$ & $(-4042,i1391)$ & $(4998,-i1872)$  \\
       \hline\hline
      \multicolumn{5}{c}{$(1679,-i118)$ [spin=1]}\\\hline
       &  $K^*\bar{K}^*$ & $\rho\rho$ & $\rho\omega$  & $\rho\phi$ \\\hline
 $g$ &  $(1040,-i1989)$ & $(6961,-i4585)$ & $0$ & $0$  \\
      \hline\hline
      \multicolumn{5}{c}{$(1569,-i16)$ [spin=2]}\\\hline
       &  $K^*\bar{K}^*$ & $\rho\rho$ & $\rho\omega$  & $\rho\phi$ \\\hline
 $g$ &  $(10208,-i337)$ & $0$& $(-4598,i451)$ & $(6052,-i604)$  \\
 \hline\hline
    \end{tabular} 
       \end{table*}
\begin{table*}[htpb]
      \renewcommand{\arraystretch}{1.5}
     \setlength{\tabcolsep}{0.1cm}
     \centering
     \caption{The same as Table \ref{table:res1}, but for the strangeness=1 and isospin=1/2 channel.\label{table:res3}}
     \begin{tabular}{c|ccc}
     \hline\hline
      \multicolumn{4}{c}{$(1643,-i24)$ [spin=0]}\\\hline
       &  $\rho K^*$ & $K^*\omega$ & $K^*\phi$    \\\hline
 $g$ &  $(8102,-i959)$ & $(1370,-i146)$ & $(-1518,i209)$  \\
  \hline\hline
      \multicolumn{4}{c}{$(1737,-i82)$ [spin=1]}\\\hline
       &  $\rho K^*$ & $K^*\omega$ & $K^*\phi$    \\\hline
 $g$ &  $(7261,-i3284)$ & $(1529,-i1307)$ & $(-1388,i1721)$  \\
   \hline\hline
      \multicolumn{4}{c}{$(1431,-i1)$ [spin=2]}\\\hline
       &  $\rho K^*$ & $K^*\omega$ & $K^*\phi$    \\\hline
 $g$ &  $(10901,-i71)$ & $(2267,-i13)$ & $(-2898,i17)$  \\
 \hline\hline
    \end{tabular} 
       \end{table*}
\begin{table*}[htpb]
      \renewcommand{\arraystretch}{1.5}
     \setlength{\tabcolsep}{0.3cm}
\caption{The properties, (mass, width) [in units of MeV], of the 11  dynamically
generated states and, if existing, of those of their PDG
counterparts. Theoretical masses and widths are obtained from two
different ways: ``pole position'' denotes the numbers obtained from
the pole position on the complex plane, where the mass corresponds to the
real part of the pole position and the width corresponds to 2 times
the imaginary part of the pole position (the box diagrams
corresponding to decays into two pseudoscalars are not included);
"real axis" denotes the results obtained from real axis amplitudes
squared, where the mass corresponds to the energy at which the amplitude
squared has a maximum and the width corresponds to the difference
between the two energies, where the amplitude squared is half of the
maximum value. (In this case, the box amplitudes corresponding to
decays into two pseudoscalars are included). The two entries under ``real axis''
are obtained with different $\Lambda_b$ as explained in the main text.
\label{table:sum}}
\begin{center}
\begin{tabular}{c|c|cc|ccc}\hline\hline
$I^{G}(J^{PC})$&\multicolumn{3}{c|}{Theory} & \multicolumn{3}{c}{PDG data}\\\hline
              & Pole position &\multicolumn{2}{c|}{Real axis} & Name & Mass & Width  \\
              &               & $\Lambda_b=1.4$ GeV & $\Lambda_b=1.5$ GeV &           \\\hline
$0^+(0^{++})$ & (1512,51) & (1523,257) & (1517,396)& $f_0(1370)$ & 1200$\sim$1500 & 200$\sim$500\\
$0^+(0^{++})$ & (1726,28) & (1721,133) & (1717,151)& $f_0(1710)$ & $1724\pm7$ & $137\pm 8$\\
$0^-(1^{+-})$ & (1802,78) & \multicolumn{2}{c|} {(1802,49)}   & $h_1$\\
$0^+(2^{++})$ & (1275,2) & (1276,97) & (1275,111) & $f_2(1270)$ & $1275.1\pm1.2$ & $185.0^{+2.9}_{-2.4}$\\
$0^+(2^{++})$ & (1525,6) & (1525,45) &(1525,51) &$f_2'(1525)$ & $1525\pm5$ & $73^{+6}_{-5}$\\\hline
$1^-(0^{++})$    & (1780,133) & (1777,148) &(1777,172) & $a_0$\\
$1^+(1^{+-})$    & (1679,235) & \multicolumn{2}{c|}{(1703,188)} & $b_1$ \\
$1^-(2^{++})$    &  (1569,32) & (1567,47) & (1566,51)& $a_2(1700)??$
\\\hline
$1/2(0^+)$       &  (1643,47) & (1639,139) &(1637,162)&  $K_0^*$ \\
$1/2(1^+)$       & (1737,165) &  \multicolumn{2}{c|}{(1743,126)} & $K_1(1650)?$\\
$1/2(2^+)$       &  (1431,1) &(1431,56) & (1431,63) &$K_2^*(1430)$ & $1429\pm 1.4$ & $104\pm4$\\
 \hline\hline
    \end{tabular}
\end{center}
\end{table*}

In the following, we present our results channel by channel and
compare with available data. We plot results for $|T|^2$ for
different amplitudes and, in addition, we calculate the pole position
and residues of the pole, which are presented in Tables
\ref{table:res1}$\sim$\ref{table:res3}. In the absence of the box
diagrams, one can easily go to the complex plane. Around the pole
position, the amplitude can be approximated by
\begin{equation}
T_{ij}=\frac{g_i g_j}{s-s_\mathrm{pole}},
\end{equation}
where $g_i$ ($g_j$) are the couplings to channel $i$ ($j$).

The resonance parameters can be obtained  from both the pole positions on
the complex plane
and the amplitudes squared on the real axis, as explained in 
the caption of Table \ref{table:sum}. In Table
\ref{table:sum}, we summarize the resonance parameters for the
dynamically generated states obtained both ways. Available data~\cite{Amsler:2008zz} are
also given for comparison. All the results including the box diagrams shown
in this paper are
calculated with $\Lambda_b=1.4$ GeV [see Eqs.~(\ref{eq:form1},\ref{eq:form2})], unless otherwise stated. On the other hand,
in Table \ref{table:sum}, we also provide the resonance parameters calculated
with $\Lambda_b=1.5$ GeV. The comparison with those calculated with $\Lambda_b=1.4$ GeV
serves to quantify the uncertainties inherent in the calculation of the box diagrams,
which provides a mechanism for the resonances to decay into two pseudoscalars.
\subsection{Strangeness=0 and Isospin=0}
\begin{figure*}[t]
\includegraphics[scale=0.7]{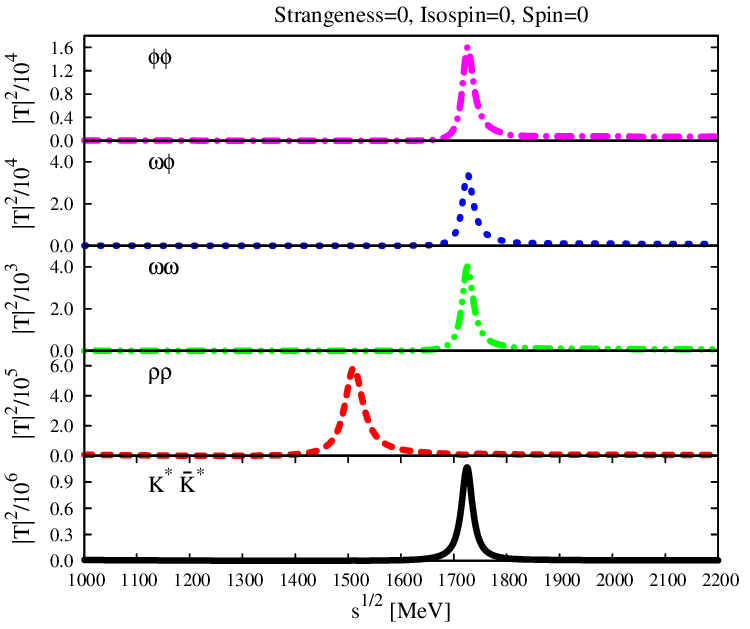} %
\includegraphics[scale=0.7]{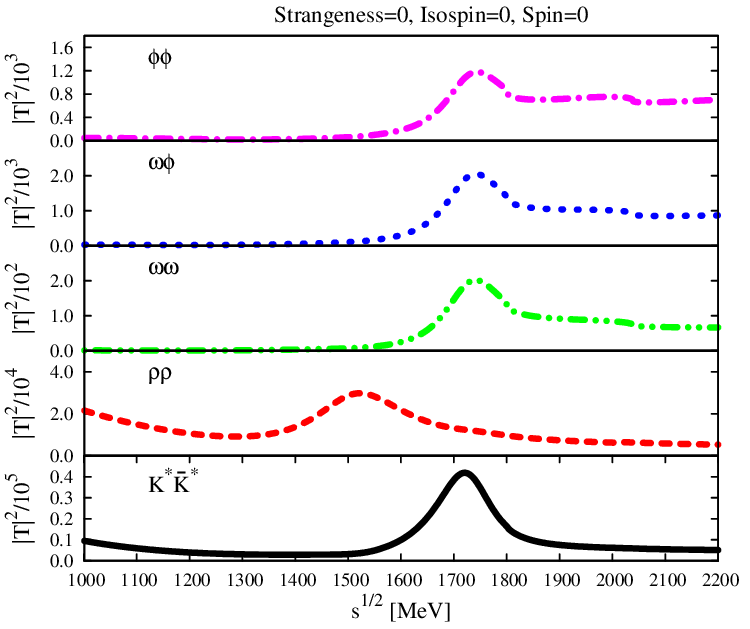}
\includegraphics[scale=0.7]{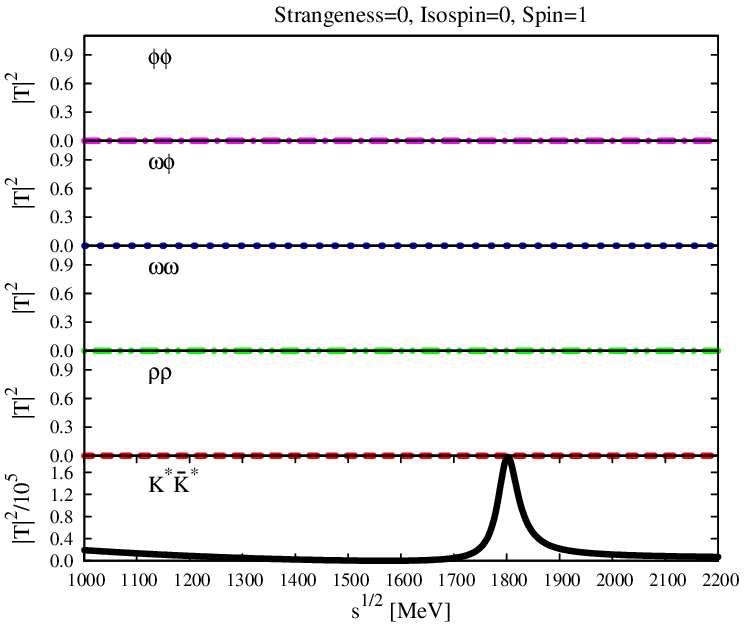}
\includegraphics[scale=0.7]{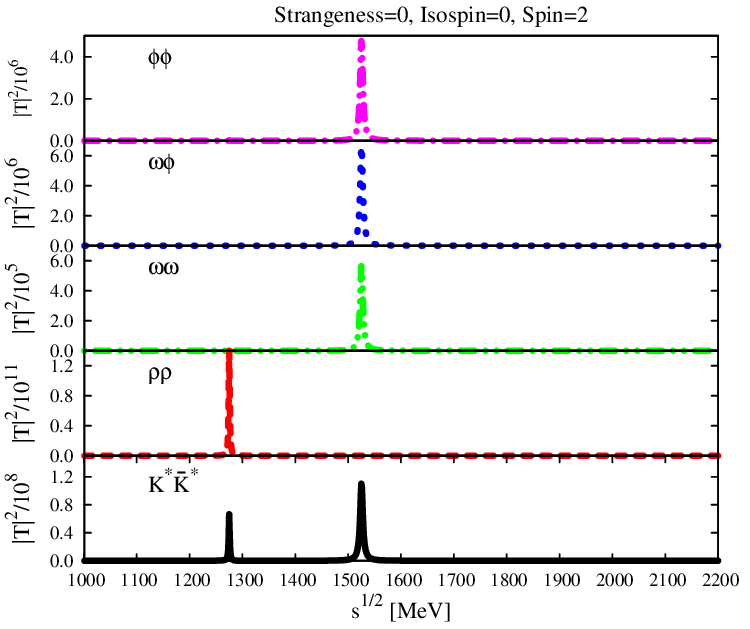}%
\includegraphics[scale=0.7]{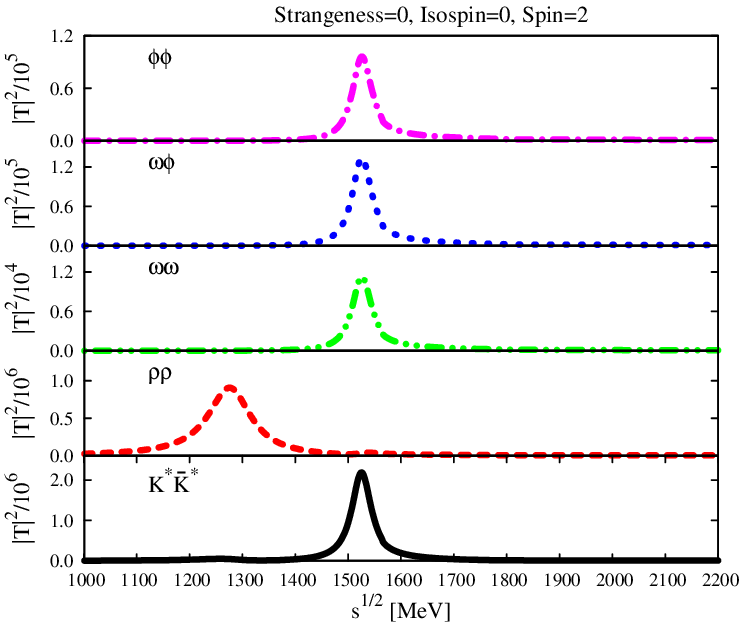}
\caption{(Color online) $|T|^2$ in $(s,I)=(0,0)$ for different spin channels without (left panel) and
with the box diagrams (right panel).\label{fig:t2one}}
\end{figure*}

In Fig.~\ref{fig:t2one}, all the $|T_{ii}|^2$'s for the
strangeness=0 and isospin=0 channel are shown
 as a function of the invariant mass of the vector-vector pair.
 The upper, middle, and bottom panels show the results for spin=0,
 spin=1, and spin=2 channels. Since the box diagrams only contribute
 to spin=0 and spin=2 channels, there are two plots in each panel for these spin channels.
 The left one shows the results without including the box diagrams, while
 the right one shows the results including the box diagrams. The comparison
 gives us an idea of the partial decay widths of the dynamically generated resonances decaying into
 two pseudoscalars. It should be noted that
 because we only consider the imaginary parts of the box diagrams,
 the pole positions on the real axis are almost the same in the two plots.

\subsubsection{Spin=0; $0^+(0^{++})$}
Two poles are found in this channel: one at $(1512,-i
26)$ MeV and another at $(1726,-i 14)$ MeV, which we associate to the
states $f_0(1370)$ and $f_0(1710)$ for the reasons given below. The
couplings of these two states to the different coupled channels
indicate that the $f_0(1370)$ is mainly a $\rho\rho$ state, while the
$f_0(1710)$ is mainly a $K^*\bar{K}^*$ state.

From the plots with the contributions of the box diagrams, the peak positions and the widths
are estimated to be (1523,257) MeV and (1721,133) MeV with the numbers in the parenthesis
being (mass, width) respectively.

The relevant information from the PDG~\cite{Amsler:2008zz} is summarized in the following:

\begin{itemize}
\item The $f_0(1370)$ has a mass of $1200\sim1500$ MeV and a width of $200\sim500$
MeV. The debate about its mass continues nowadays; while a recent
analysis advocates a mass around 1370 MeV~\cite{Bugg:2007ja}, preliminary
results from the Belle Collaboration rather point to a value around
1470 MeV~\cite{uehara}. Among its decay modes, according to the PDG~\cite{Amsler:2008zz},
the $4\pi$ mode
is larger than $72\%$, where the $\rho\rho$ mode is dominant. 
In our approach, the $\pi\pi$ mode is dominant, as can be seen from 
Table \ref{table:sum}, which is consistent with the results of 
Ref.~\cite{Albaladejo:2008qa}
and the recent analysis of D. V. Bugg~\cite{Bugg:2007ja}.

\item The $f_0(1710)$ has a mass of $1724\pm 7$ MeV and
a width of  $137 \pm 8$ MeV. The main decay channel is
through $K\bar{K}$, $\eta\eta$, and $\pi\pi$. The decay mode to
$\omega\omega$ has been seen. This is in agreement with our findings
since the two pseudoscalar box diagrams contain these decay
channels. Indeed, we find that the $K\bar{K}$ decay channel is
dominant. More specifically, our calculated branching ratios are
$\sim 55\%$ for $K\bar{K}$, $\sim27\%$ for $\eta\eta$, $<1\%$ for
$\pi\pi$, and $\sim18\%$ for the vector-vector component.  
On the other hand, the PDG gives the following
averages: $\Gamma(\pi\pi)/ \Gamma(K\bar{K}) =0.41_{-0.17}^{+ 0.11}$,
and $\Gamma(\eta\eta)/\Gamma(K \bar{K})=0.48 \pm0.15$~\cite{Amsler:2008zz}.  Our calculated branching ratio
for the $\eta\eta$ channel is in agreement with their average, while the ratio for
the $\pi\pi$ channel is much smaller. However, we notice that
the above PDG $\Gamma(\pi\pi)/ \Gamma(K\bar{K})$ ratio is taken from the BES experiment
$J/\psi\rightarrow \gamma\pi^+\pi^-$~\cite{Ablikim:2006db}, which comes from
a partial wave analysis that includes seven resonances. On the other hand,
there is another BES experiment $J/\psi\rightarrow\omega K^+K^-$~\cite{Ablikim:2004st},
which filters $I=0$ automatically and gives an upper limit
$\Gamma(\pi\pi)/ \Gamma(K\bar{K})<11\%$ at the $95\%$ confidence level. Clearly more analysis is
advised to settle the issue.

\item We see that the $f_0(1370)$ is mainly $\rho\rho$, and the
$f_0(1710)$ is mostly $K^*\bar{K}^*$. Although our picture for
the resonances would correspond, in terms of quark degrees of freedom,
to a four quark ($qq\bar{q}\bar{q}$) system, it is anyway interesting to recall
that pictures for these resonances
in terms of $q\bar{q}$ also advocate $ud$ components for the
$f_0(1370)$ and strange quark components for the $f_0(1710)$~\cite{Amsler:2008zz}.

\end{itemize}

The $f_0(1500)$, on the hand, has a mass of  $1505\pm6$ MeV and a
width of $109\pm7$ MeV. The width of the $f_0(1500)$ is too
small to be associated to the lower scalar state that we get
dynamically generated in the unitary approach, with a width of about
260 MeV.

\subsubsection{Spin=1; $0^-(1^{+-})$}
One  pole at $(1802,-i 39)$ MeV is found. However, this state
cannot be clearly identified with any of the $h_1$ states listed in
the PDG. Note that this state is built only from $K^*\bar{K}^*$. The
fact that this state couples only to $K^*\bar{K}^*$ and not to two
pseudoscalars, as we discussed for the spin=1 states, makes its
observation difficult. However, the prediction is neat; $|T|^2$ is
sizable compared to other resonances and we find a clear pole on the
complex plane associated to this resonance. On the other hand, the
energy is such that it is slightly above the $K^*\bar{K}^*$
threshold. This fact, in addition to the width of the $K^*$, would
make the observation of this state possible by looking at the
$K\bar{K}\pi\pi$ decay channel, and even the $K\pi$ resonant shape
could be partly reconstructed to give support to the $K^*\bar{K}^*$
nature of this resonance. 

\subsubsection{Spin=2; $0^+(2^{++})$}
Two poles are found on the complex plane: one at $(1275,-i
1)$ MeV and the other at $(1525,-i 3)$ MeV, which we associate to
$f_2(1270)$ and $f'_2(1525)$. The lower one mainly couples to
$\rho\rho$ and very weakly to $K^*\bar{K}^*$. This can be seen in
the strengths of $|T|^2$ in the lower rightmost panel of
Fig.~\ref{fig:t2one}, and more clearly in the value of $g$ for the
couplings to the channels as shown in Table \ref{table:res1}. The
higher resonance couples mainly to $K^*\bar{K}^*$, $\omega\phi$, and
$\phi\phi$. As mentioned above, the masses of these two states have been
used to fine-tune our subtraction constants.

From $|T|^2$ on the real axis obtained including box diagrams, one
obtains the masses and widths as $(1276,97)$ MeV and $(1525,45)$ MeV. 
It is gratifying to see that the estimated widths
are smaller than their experimental counterparts [$185.0^{+2.9}_{-2.4}$ MeV for
the $f_2(1270)$ and $73^{+6}_{-5}$ MeV for the $f'_2(1525)$]. This should always
be the case since other coupled channels, which we
have not included, may also contribute. However, note that the order of
magnitude is consistent and furthermore we predict a bigger width
for the $f_2(1270)$ than for the $f'_2(1525)$ in spite of the fact
that the higher mass resonance has more phase space to decay.
We also see in Table \ref{table:sum} that these widths get a bit bigger
by increasing moderately the value of the $\Lambda_b$ parameter of the form factors
of Eqs.~(\ref{eq:form1},\ref{eq:form2}).

Once again it is interesting to compare the partial decay widths. For
the $f_2(1270)$ we get most of the width from $\pi\pi$ decay. In the
PDG the branching ratios are $84.8\%$ for $\pi\pi$, $4.6\%$ for
$KK$, and $<1\%$ for $\eta\eta$~\cite{Amsler:2008zz}, to be compared with our calculated numbers $\sim88\%$ for $\pi\pi$, $\sim10\%$ for $K\bar{K}$ and $<1\%$ for $\eta\eta$.

The case of the $f'_2(1525)$ is equally clarifying. We get
most of the width from $K\bar{K}$ ($\sim66\%$, compared to the branching
ratio of $88.7\%$ in the PDG~\cite{Amsler:2008zz}). Our calculated ratios are 
$\sim 21\%$ for $\eta\eta$, $\sim1\%$ for $\pi\pi$, 
and $\sim13\%$ for the vector-vector component,
while the PDG gives $10.4\%$ for $\eta\eta$ and $0.8\%$ for $\pi\pi$~\cite{Amsler:2008zz}.
The agreement  is reasonable.

The position of the higher state at 1525 MeV is also close to the $f_2(1430)$
and the $f_2(1565)$. The $f_2(1430)$ is a little further away while
the $f_2(1565)$ has a strong coupling to $\rho\rho$ decay mode,
while in our calculation this state couples very weakly to the
$\rho\rho$ channel; therefore, we do not favor the assignment
to any of these two resonances.

 \begin{figure*}[htpb]
\includegraphics[scale=0.7]{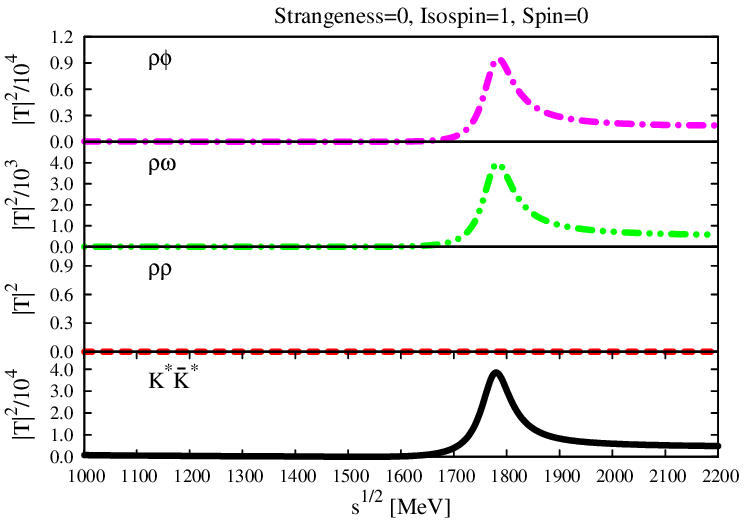}%
\includegraphics[scale=0.7]{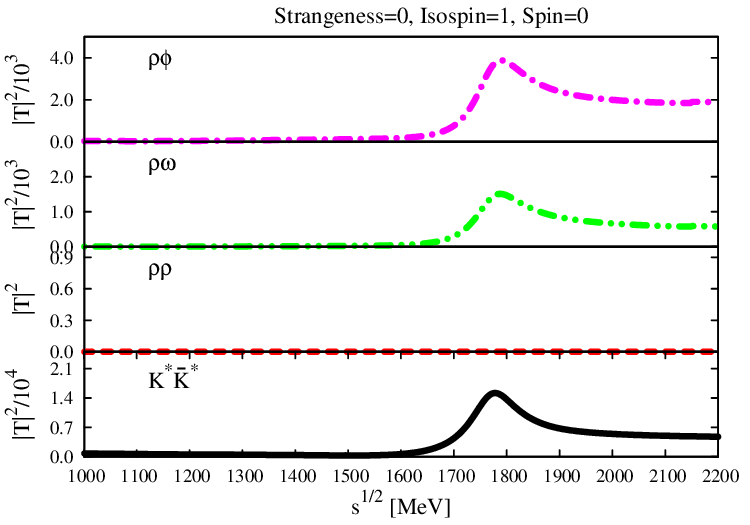}
\includegraphics[scale=0.7]{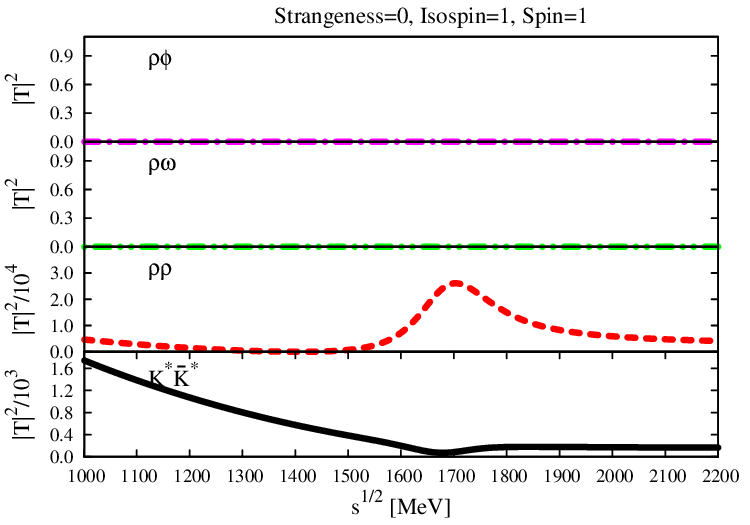}
\includegraphics[scale=0.7]{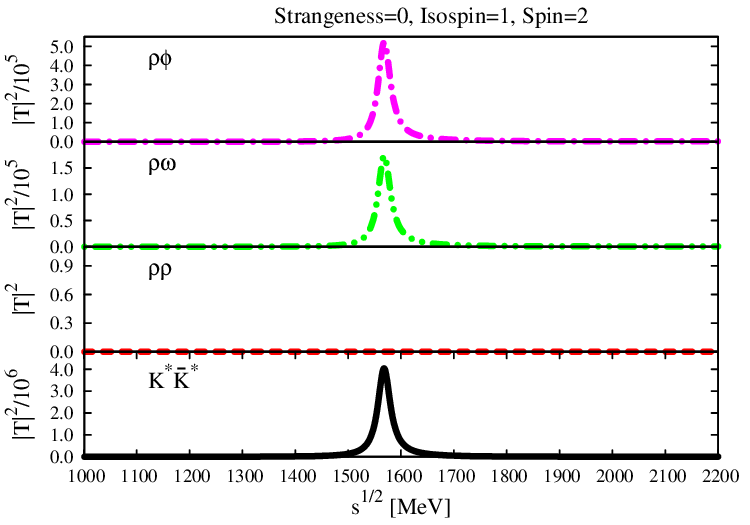}%
\includegraphics[scale=0.7]{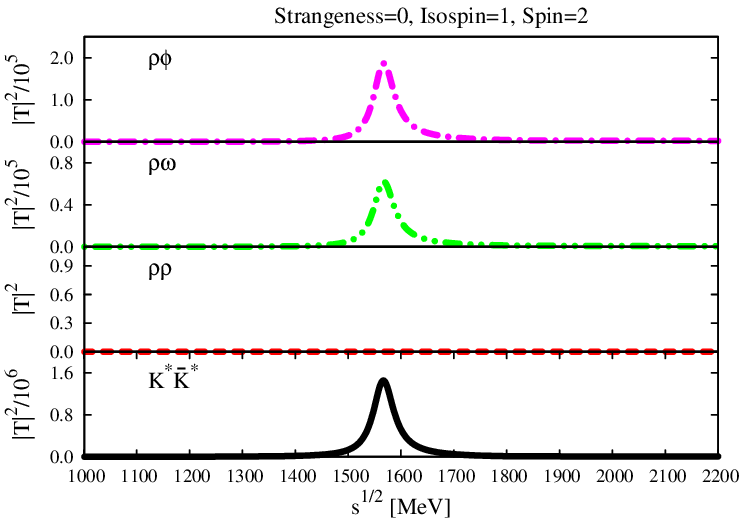}
\caption{(Color online) The same as Fig.~\ref{fig:t2one}, but for the $(s,I)=(0,1)$ channel.\label{fig:s0i1}}
\end{figure*}

\begin{figure*}[htpb]
\includegraphics[scale=0.7]{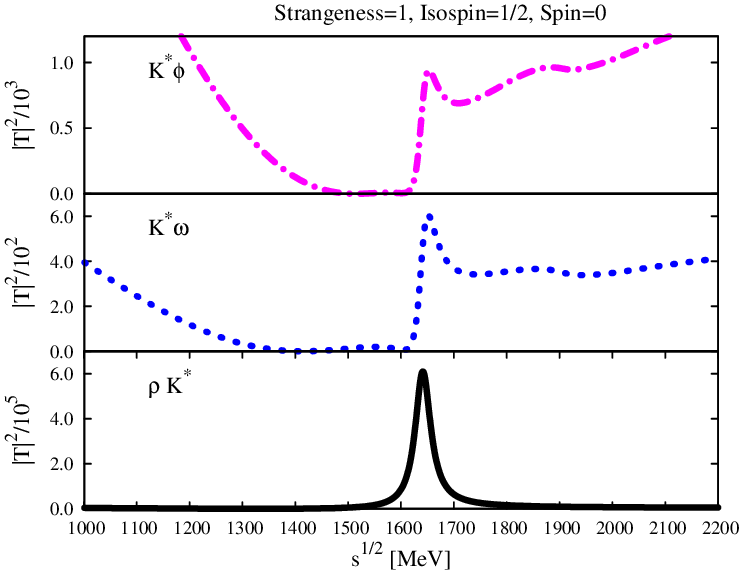}%
\includegraphics[scale=0.7]{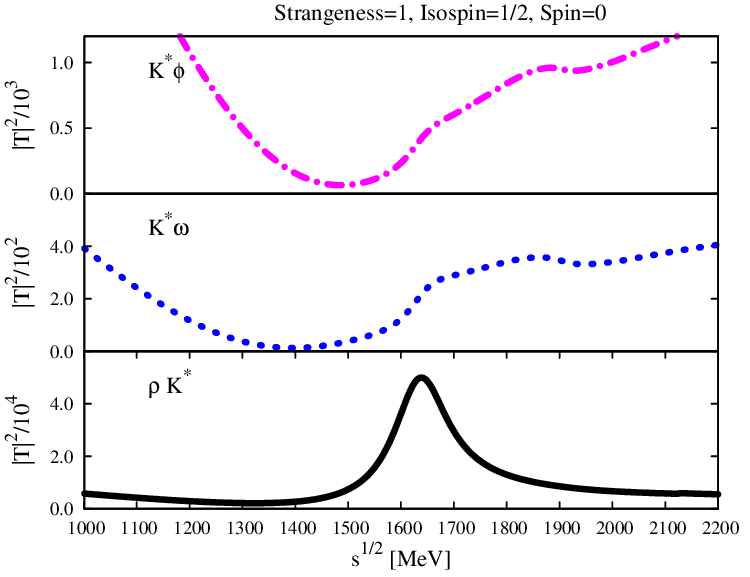}
\includegraphics[scale=0.7]{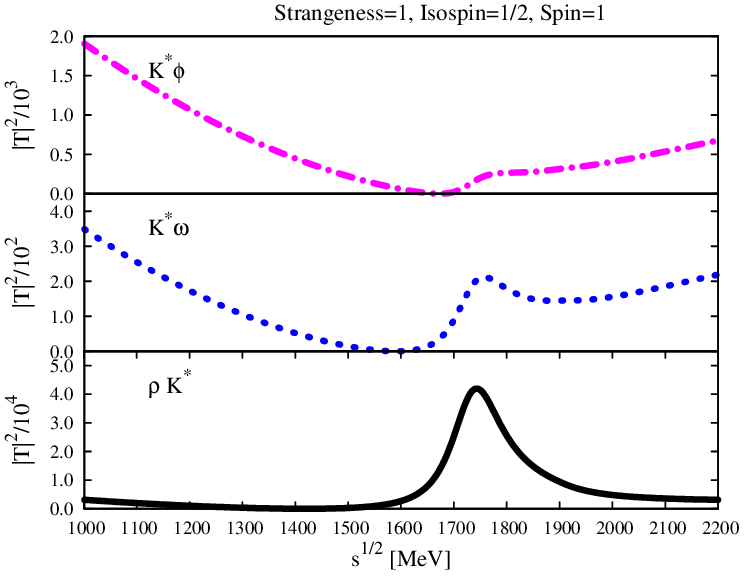}
\includegraphics[scale=0.7]{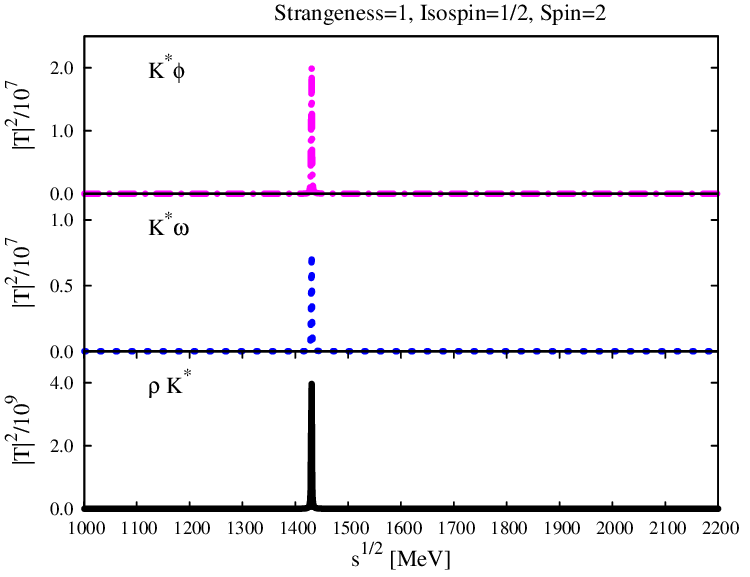}%
\includegraphics[scale=0.7]{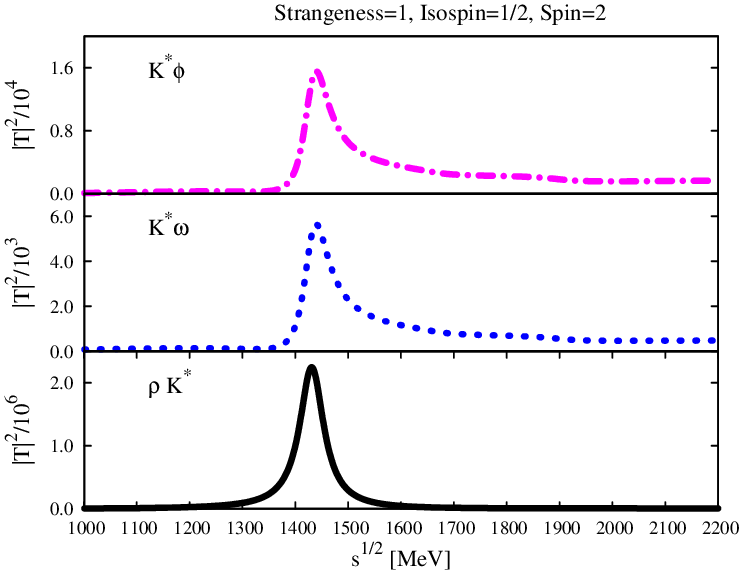}
\caption{(Color online) The same as Fig.~\ref{fig:t2one}, but for the $(s,I)=(1,1/2)$ channel.\label{fig:s1i1h}}
\end{figure*}

\subsection{Strangeness=0 and Isospin=1}
In Fig.~\ref{fig:s0i1}, we plot $|T_{ii}|$'s for the
strangeness=0 and isospin=1 channel. Three resonances are
found dynamically generated.
\subsubsection{Spin=0; $1^-(0^{++})$}
One pole is found at $(1780,-i 66)$ MeV, and it has the quantum
numbers of $a_0$. It couples mostly to the $K^*\bar{K}^*$ channel. No
$a_0$ around this energy region has been reported, according to the
PDG~\cite{Amsler:2008zz}.

Including the box diagrams, one gets $(1777,148)$ MeV. It is seen that the inclusion
of the box diagrams does not change much both the mass and the width of this state,
meaning that it has a small branching ratio to two pseudoscalars.

 This resonance can in principle be formed in
 $J/\psi\rightarrow \gamma K^*\bar{K}^*$ and
$J/\psi\rightarrow \gamma K\bar{K}$. It is below the $K^*\bar{K}^*$ threshold
and wider than the $0^-(1^{+-})$ state, and it could produce a
broader bump close to the $K^*\bar{K}^*$ threshold. Such a feature
does seem to show up in the BES experiment~\cite{Bai:1999mk}, but once again a new
look at these data would be worthwhile.

\subsubsection{Spin=1; $1^+(1^{+-})$}
This channel has the quantum numbers of $b_1$. One pole is found  at
$(1679,-i 118)$ MeV, and it couples strongly to the $\rho\rho$
channel. Experimentally, no $b_1$ has been reported around this energy
region.

This state does not decay into $\pi\pi$ but there
should be no problem in studying the $\rho\rho$ invariant mass since
the mass of the particle appears above the $\rho\rho$ threshold.
Although several experiments have looked into $J/\psi\rightarrow
2(\pi^+\pi^-)\pi^0$ ~\cite{JeanMarie:1975gg,Burmester:1977ra,Franklin:1983ve,Augustin:1988ja}, none
of them has looked at the $\rho\rho$ invariant mass distribution. We
can only encourage further search in this direction once the
previous works have proved the viability of the experiment.

\subsubsection{Spin=2; $1^-(2^{++})$}
One pole is found at $(1569,-i 16)$ MeV, and it couples strongly to
$K^*\bar{K}^*$. Including the box diagrams, one
obtains $(1567,47)$ MeV. The closest $a_2$ in energy included in the PDG is
the $a_2(1700)$ with a mass of $1732\pm16$ MeV and  a width of
$194\pm40$ MeV, whose decay to $\omega\rho$ has been seen~\cite{Amsler:2008zz}. It should be
noted that the properties of this particle are not well determined. Different
experiments report quite different values for both its mass and width~\cite{Amsler:2008zz}.

In order to see if the resonance we get could be associated to the
$a_2(1700)$, we have changed the values of the subtraction constants to
move its pole position to larger mass values. For instance, if we change
the value of $a_{K^*\bar{K^*}}$ from $-1.726$ [determined by the
$f'_2(1525)$ mass] to $-1.0$, we would have a mass of 1704 MeV and a width of 49 MeV.
The mass would be much closer to the PDG average but the width would still be much smaller.
A modification of the values of the subtraction constants of the other two coupled channels
($\rho\omega$ and $\rho\phi$)
leads to similar conclusions. Given the large uncertainty
in the experimental status of the $a_2(1700)$, we find no particular reason to
associate the state we find dynamically to this resonance.

We also note that 
the modification of $a_{K^*\bar{K}^*}$  
has small influences on the states with the quantum numbers
of $b_1$ and $a_0$, which we studied in the two preceding subsections, and it does not allow us to associate these two states with 
any well-known resonances listed in the PDG.  
\subsection{Strangeness=1 and Isospin=1/2}
In Fig.~\ref{fig:s1i1h}, we plot $|T_{ii}|^2$'s for the
strangeness=1 and isospin=1/2 channel.
\subsubsection{Spin=0; $1/2(0^+)$}
One pole is found at $(1643,-i 24)$ MeV, and it couples strongly to
$\rho K^*$. Including the box diagrams, one obtains $(1639,139)$ MeV.

At first sight,  this state might be the $K(1630)$. On the other
hand, the $K(1630)$ [$1/2(?^?)$],  with a mass of $1629\pm7$ MeV and
a width of $16^{+19}_{-16}$ MeV~\cite{Amsler:2008zz}, might be too narrow to be
associated with the state dynamically generated from vector-vector
interaction. There is another indication not to associate the state
we find with the $K(1630)$, since our main decay mode is $\pi K$ from
the two meson box diagrams, while the decay mode observed in the PDG
is  $K\pi^+\pi^-$.

\begin{figure*}[htpb]
\includegraphics[scale=0.7]{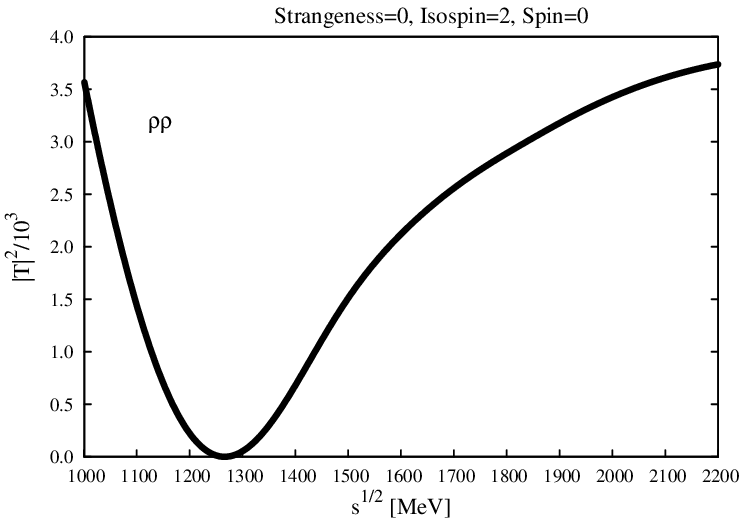}%
\includegraphics[scale=0.7]{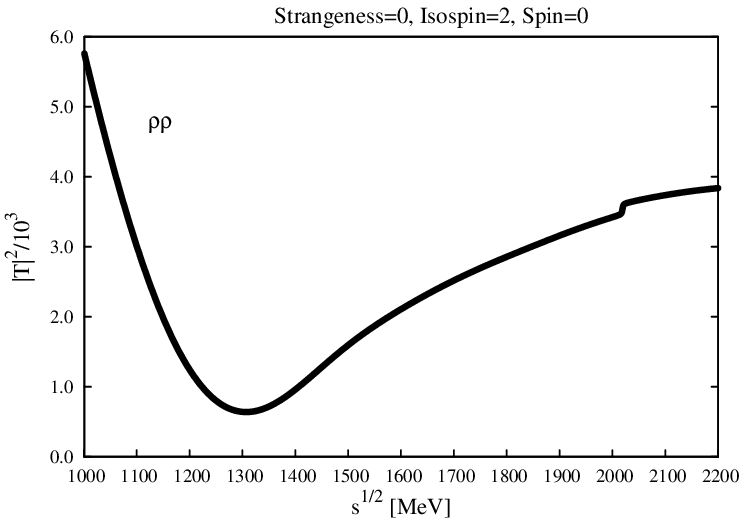}
\includegraphics[scale=0.7]{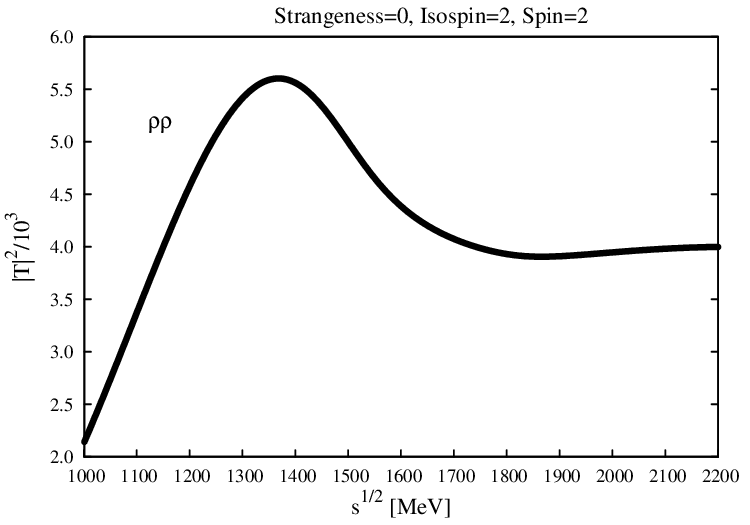}%
\includegraphics[scale=0.7]{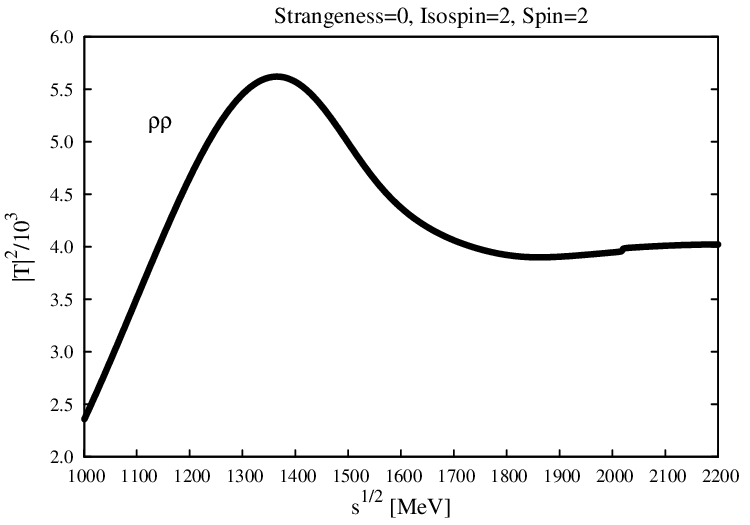}
\caption{The same as Fig.~\ref{fig:t2one}, but for the $(s,I)=(0,2)$ channel.
Note that we have not shown the results for spin=1 channel, since there are 
no interactions here
because of the properties of identical particles. \label{fig:s0i2}}
\end{figure*}
\begin{figure*}[htpb]
\includegraphics[scale=0.7]{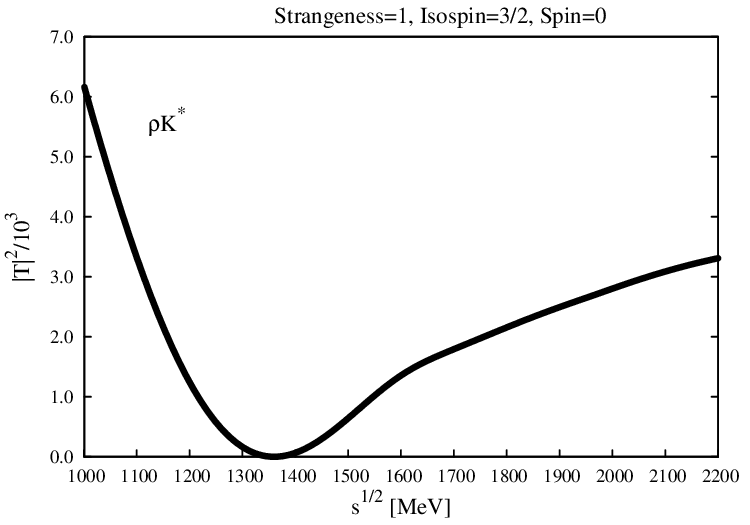}%
\includegraphics[scale=0.7]{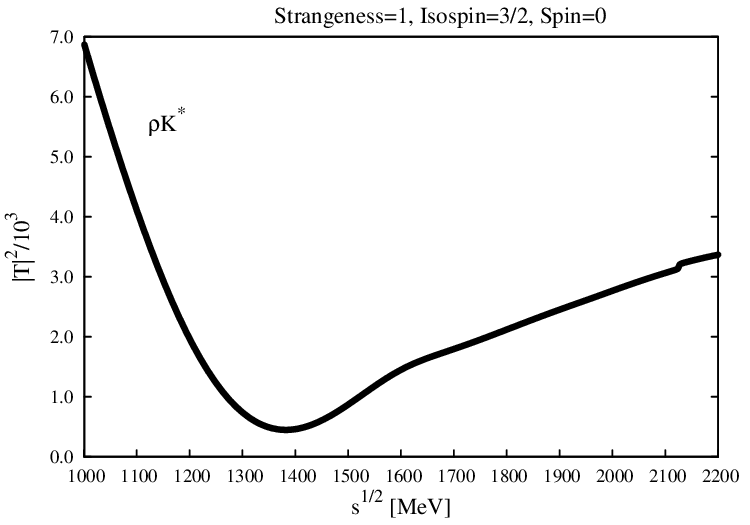}
\includegraphics[scale=0.7]{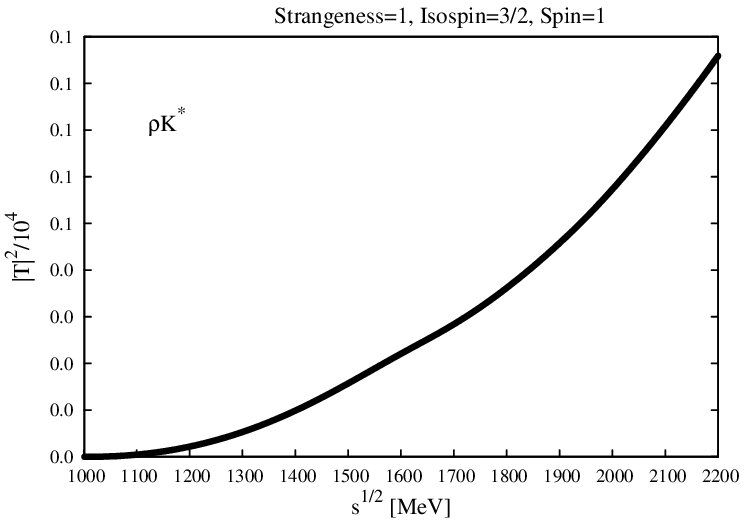}
\includegraphics[scale=0.7]{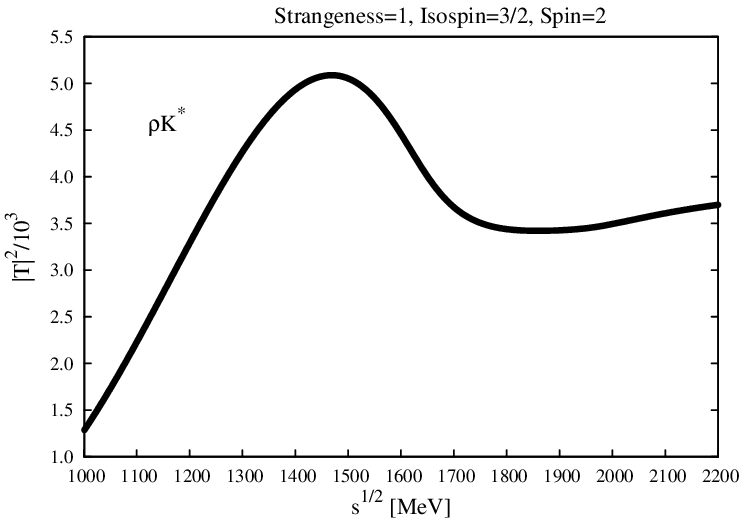}%
\includegraphics[scale=0.7]{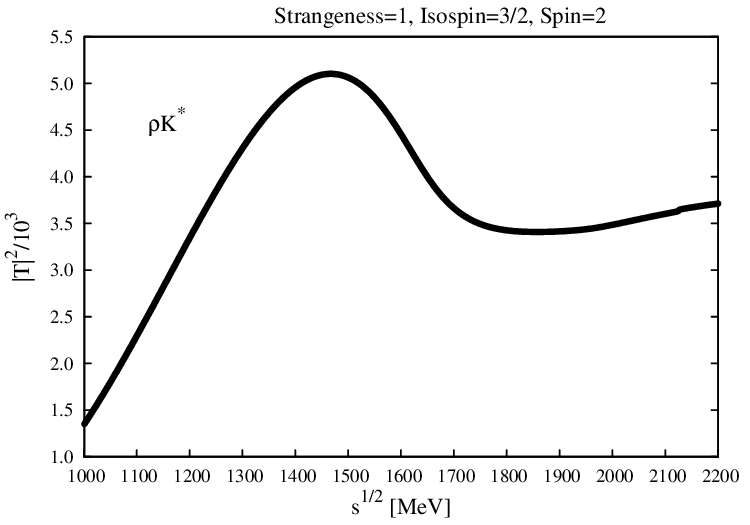}
\caption{The same as Fig.~\ref{fig:t2one}, but for the $(s,I)=(1,3/2)$ channel.\label{fig:s1i3h}}
\end{figure*}

\begin{figure*}[htpb]
\includegraphics[scale=0.7]{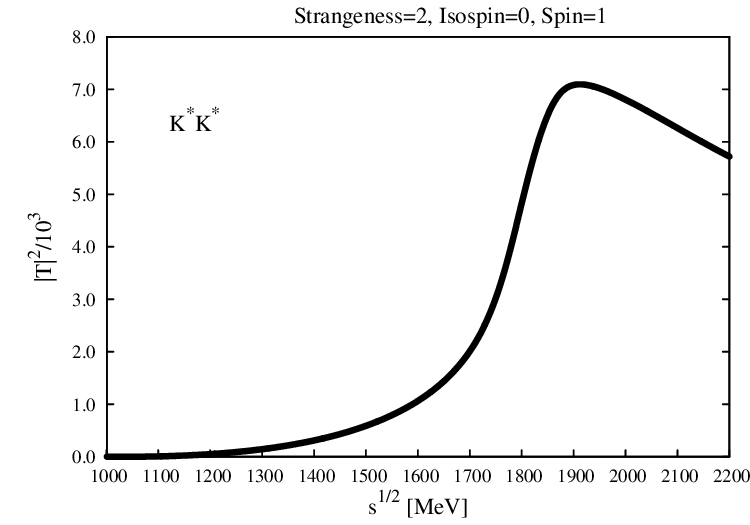}
\caption{The same as Fig.~\ref{fig:t2one}, but for the
$(s,I)=(2,0)$ channel. There are no interactions in spin=0 and spin=2 channels
due to the properties of identical particles.  \label{fig:s2i0}}
\end{figure*}

\begin{figure*}[htpb]
\includegraphics[scale=0.7]{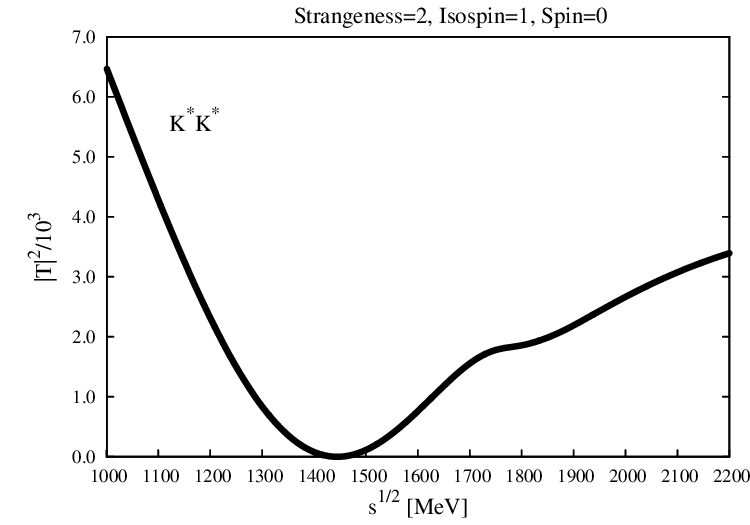}%
\includegraphics[scale=0.7]{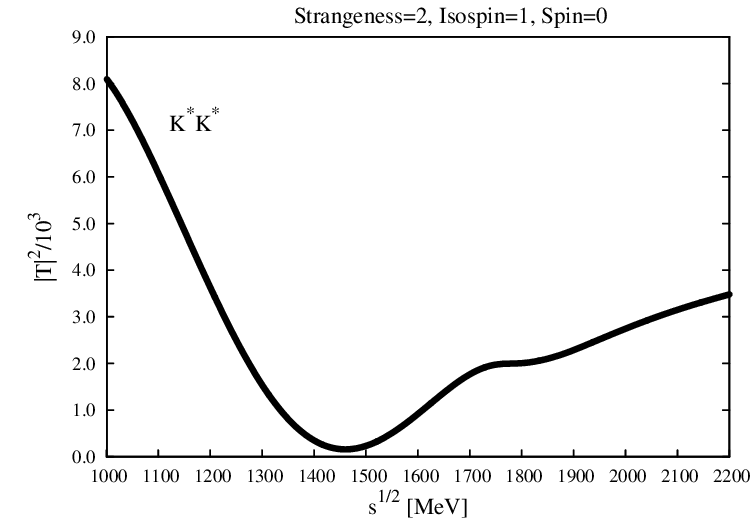}
\includegraphics[scale=0.7]{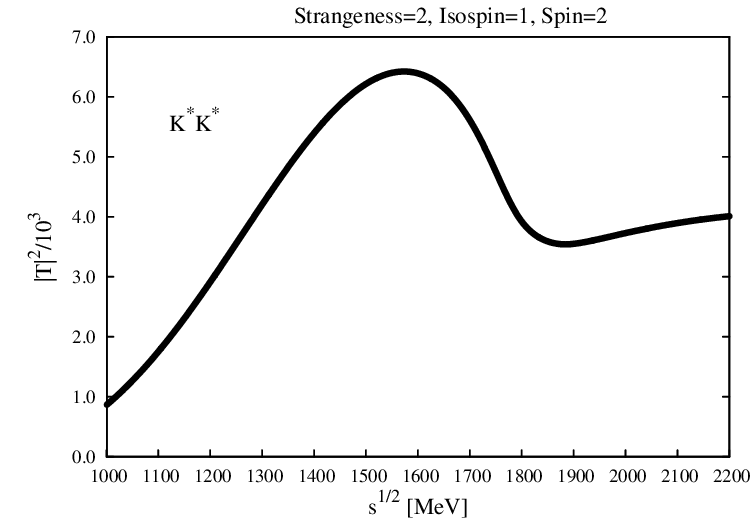}%
\includegraphics[scale=0.7]{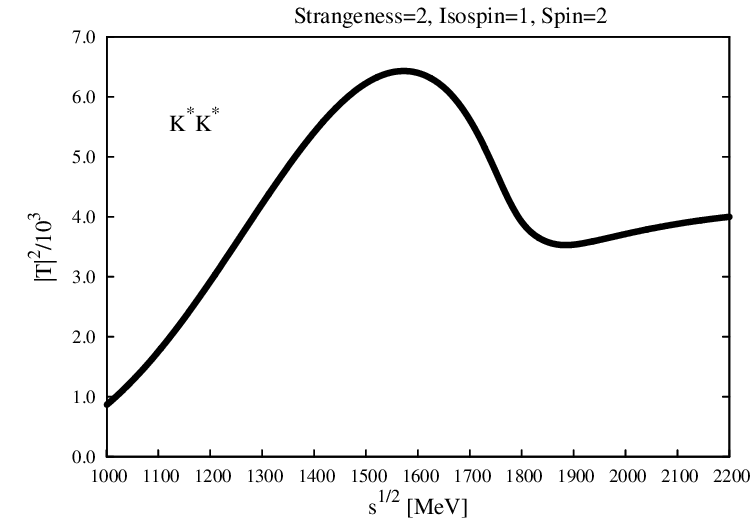}
\caption{The same as Fig.~\ref{fig:t2one}, but for the
$(s,I)=(2,1)$ channel. There are no interactions in spin=1 channel due to the properties of
identical particles.\label{fig:s2i1}}
\end{figure*}
\subsubsection{Spin=1; $1/2(1^+)$}
One pole is found  at $(1737,-i 82)$ MeV, and it couples strongly
to $\rho K^*$. No $K_1$ around this energy region is reported in the
PDG, with the closest one being the $K_1(1650)$ with a mass of
$1650\pm50$ MeV and a width of $150\pm50$ MeV~\cite{Amsler:2008zz}. The width of the
$K_1(1650)$ is 150 MeV, and we also obtain a width of about 160 MeV.
Since the width is twice as large as the difference of masses the
association of these two states is tempting. There is another
feature that could support this association; in spite of the limited information
on this resonance, the only decay channels observed are $K\pi\pi$,
$K\phi$, but none on two pseudoscalars for which there is more phase
space. This is in agreement with the fact that 
our state of spin 1 does not decay into two pseudoscalars, as
we have mentioned.
\subsubsection{Spin=2; $1/2(2^+)$}
One pole is found at $(1431,-i 1)$ MeV, which might correspond to
the $K^*_2(1430)$, and its position has been used to fine-tune the
subtraction constants in this channel. Including the box diagrams,
one obtains $(1431,56)$ MeV.

According to the PDG, the $K^*_2(1430)$ has a mass of $1429\pm1.4$
MeV and a width of $104\pm4$ MeV. Among its decays modes, the
$K^*\pi\pi$ mode amounts to $(13.4\pm2.2)\%$, some of which might be
$\rho K^*$; the $K \pi$ mode amounts to $(49.9\pm 1.2)\%$.
Therefore, the width that we obtain is in reasonable agreement
with the data.

\subsection{Other channels}
It is interesting to note that out of the 21 combinations of
strangeness, isospin and spin, we have found resonances only in nine of
them. In all the ``exotic'' channels, from the point of view that they cannot be formed from $q\bar{q}$
states, we did not find dynamically
generated resonances, including the three strangeness=0, isospin=2
channels, the three strangeness=1, isospin=3/2 channels, the six
strangeness=2 channels (with either isospin=0 or isospin=1).

It is also interesting to note that although no poles are found on
the complex plane, there do exist some structures on the real axis.
For instance, in the (strangeness=0, isospin=2) channel, one finds a
dip around $\sqrt{s}=1300$ MeV in the spin=0 channel, and a broad
bump in the spin=2 channel around $\sqrt{s}=1400$ MeV, as can be
clearly seen from Fig.~\ref{fig:s0i2}. In the
(strangeness=1, isospin=3/2) and (strangeness=2, isospin=1) channels,
one observes similar structures occurring at shifted energies due to
the different masses of the $\rho$ and the $K^*$, as can be seen
from Figs.~\ref{fig:s1i3h} and \ref{fig:s2i1}.

It is worthwhile mentioning that we obtain some broad bumps in the
following four channels: (strangeness=0, isospin=2, spin=2),
 (strangeness=1, isospin=3/2, spin=2),  (strangeness=2, isospin=0, spin=1),
and (strangeness=2, isospin=1, spin=2),
see Figs.~7-10. All these are exotic channels. As mentioned before, none of the broad peaks corresponds to a pole on the complex plane, and hence, according to the common criteria, they do
not qualify as resonances. Let us see what is the experimental
information in these sectors. In the PDG~\cite{Amsler:2008zz}, we find the $X(1600)$ with
strangeness=0 and quantum numbers $2^+(2^{++})$ with a mass of
$1600\pm100$ MeV and a width of $400\pm 200$ MeV. There are
candidates in theoretical models for this. Indeed, based on a theoretical estimate
of the twist 4 contributions in explaining the recent L3 data on $\gamma^*\gamma\rightarrow
\rho^0\rho^0$ and $\gamma^*\gamma\rightarrow \rho^+\rho^-$~\cite{Achard:2003qa,Achard:2004ux}, I.V. Anikin et al. advocate the existence of an exotic isotensor resonance with 
a mass of $\sim1.5$ GeV and a width of $\sim0.4$ GeV~\cite{Anikin:2005ur}.
However, we can offer here a different interpretation
for the experimental bump, since it
might be identified with the broad bump that we get with these
quantum numbers around 1400 MeV and a similar width. Indeed, the
experiment where the bump is reported~\cite{Albrecht:1990cr}  sees it in the
$\rho^0\rho^0$ channel. It looks rather clear that the bump observed
is the one we find in the $\rho\rho$ amplitude, but this does not
qualify as a resonance. 

One can also speculate about the scalar $2^+(0^{++})$ state reported
in the PDG around 1400 MeV from a weak signal found as a broad bump in
Ref.~\cite{Filippi:2000is}. As can be seen from the upper panel of Fig. \ref{fig:s0i2},
 we find a dip in $\rho\rho$ amplitude squared in this channel around 1300 MeV.
Such a dip in the $\rho\rho$ amplitude can lead to a bump in $\pi^+\pi^+$ production, in an analogous
way as  what occurs to the $f_0(980)$ resonance, which shows
up as a dip in the $\pi\pi$ cross section but as a peak in $\gamma\gamma$ or
other production processes~\cite{Oller:2000ma}. Once again, the bump could not be
associated to a pole in our approach and, hence, would not qualify as a resonance.

We do not find any candidate in the PDG to our broad bumps in the
strange sector. However,  the findings of the present work should be
kept in mind in the verge of possible claims for exotic strange
mesons from bumps observed in cross sections.

We would like to give some perspective to the results obtained here. We have used as building
blocks for our states only vector mesons. Two pseudoscalar states have been considered
for the decay but not incorporated as coupled channels. Other possible channels, like
$\sigma\sigma$ in the case of $\rho\rho$ scattering, are also omitted in our approach.
The contributions of these channels in a coupled channel approach would be advisable
should one try to get, for instance, $\pi\pi$ scattering in a broad range of energies. Such
an approach has been undertaken in Ref.~\cite{Albaladejo:2008qa}.\footnote{This work is now being extended and we do not elaborate further on it, but one should keep track
of new developments along this line~\cite{oller}.} However, this is not 
our purpose here. We take only vector mesons as building blocks with their respective interactions,
and we look at the states that are generated dynamically from these interactions. We then get a few meson
resonances, but not all. This tells us which resonances are most likely to be essentially 
vector-vector ``molecules,'' and this is the purpose of the present work.

\section{Summary and conclusions}
We have performed a study of vector meson-vector meson interaction
using a unitary approach. Employing the coupled channel
Bethe-Salpeter equation to unitarize the tree-level transition
amplitudes obtained from the hidden-gauge Lagrangians, 11 states
in nine strangeness-isospin-spin channels are dynamically
generated. Among them, five states are associated to those reported in
the PDG, i.e., the $f_0(1370)$, the $f_0(1710)$, the $f_2(1270)$,
the $f'_2(1525)$, the $K_2^*(1430)$. The association of two other
states, the $a_2(1700)$ and the $K_1(1650)$, are likely, particularly the
$K_1(1650)$, but less
certain. The $f_0(1370)$ and $f_2(1270)$ have already been reported
in Ref.~\cite{Molina:2008jw}, and they are built mainly from the
$\rho\rho$ interaction. We reconfirm the findings of this early work 
after including
all SU(3) coupled channels. The box diagrams in our approach provide
a mechanism for the dynamically generated states to decay into two
pseudoscalars. This mechanism broadens the scalar states and the
tensor states in the strangeness=0 and isospin=0 channel and the
strangeness=1 and isospin=1/2 channel but not for the spin=1 states.
On the other hand, this mechanism contributes little to the widths
of the scalar and tensor states in the strangeness=0 and isospin=1
channel.

We have used the masses of the $f_2(1270)$, the $f'_2(1525)$, and
the $K_2^*(1430)$ to fine-tune the free parameters of the approach,
the subtraction constants in the vector-vector loop functions.
After this is done, there is little freedom in changing 
the total decay widths and practically no freedom in changing the
decay branching ratios.
It is then gratifying to see that the total and partial decay widths of these
resonances are consistent with the data.  It is also interesting to
see that the two $f_0$ states appear at proper positions with
reasonable widths compared to the data. 

Four of the 11 dynamically generated states can not be
associated with known states in the PDG. These states either have
small branching ratios into two pseudoscalars or are in the
strangeness=1 sector, where the experimental situation is less
satisfactory than in the strangeness=0 sector.

Another interesting finding of our work is the broad bumps found in
four exotic channels, none of which corresponds to poles on the
complex plane. One of these bumps is identified with the structure
of the $X(1600)$, which is reported in the PDG as a resonant state
with $2^+(2^{++})$. Our study provides an interpretation of this
bump, stemming from the $\rho\rho$ interaction in this channel,
which, however, does not have any pole associated and, hence, does
not qualify as a resonance.

For the resonances predicted and not reported in the PDG we have offered
suggestions on how they could be searched experimentally with
present experiment facilities, and we can only encourage further work
in this direction.

\section{Acknowledgments}
L. S. Geng thanks R. Molina,  L. Alvarez-Ruso, and M. J. Vicente Vacas for
useful discussions. This work is partly
supported by DGICYT Contract No. FIS2006-03438 and the EU Integrated
Infrastructure Initiative Hadron Physics Project under contract
RII3-CT-2004-506078.

\section{Appendix}
\subsection{Tree-level transition amplitudes of the four-vector-contact
diagrams and of the $t$($u$)-channel vector-exchange diagrams 
for different strangeness, isospin and spin channels.}

\begin{table}[htpb]
      \renewcommand{\arraystretch}{1.5}
     \setlength{\tabcolsep}{0.3cm}
     \centering
     \caption{The $V_{ij}$'s of the four-vector-contact term in the strangeness=0, isospin=0 and spin=0 channel.\label{table:vvvv1}}
     \begin{tabular}{c|ccccc}
     \hline\hline
       &  $K^*\bar{K}^*$ & $\rho\rho$ & $\omega\omega$  & $\omega\phi$  & $\phi\phi$ \\\hline
 $K^*\bar{K}^*$ &  $6 g^2$          & $2\sqrt{3} g^2$ & $-2 g^2$ & $4 g^2$ & $-4 g^2$ \\
 $\rho\rho$          &   & $8 g^2$        & 0 & 0 & 0 \\
 $\omega\omega$      &   &  & 0          & 0 & 0 \\
 $\omega\phi$        &  &  &   & 0 & 0 \\
 $\phi\phi$          & & & & & 0\\\hline\hline
    \end{tabular} 
       \end{table}

\begin{table}[htpb]
      \renewcommand{\arraystretch}{1.5}
     \setlength{\tabcolsep}{0.2cm}
     \centering
     \caption{The same as Table \ref{table:vvvv1}, but for the
     strangeness=0, isospin=0 and spin=1 channel.\label{table:vvvv2}}
     \begin{tabular}{c|ccccc}
     \hline\hline
       &  $K^*\bar{K}^*$ & $\rho\rho$ & $\omega\omega$  & $\omega\phi$  & $\phi\phi$ \\\hline
 $K^*\bar{K}^*$ &  $9g^2 $          & 0 & 0 & 0 & 0 \\
 $\rho\rho$          &   & 0       & 0 & 0 & 0 \\
 $\omega\omega$      &   &  & 0          & 0 & 0 \\
 $\omega\phi$        &   &  &  & 0 & 0 \\
 $\phi\phi$          &   &  &  &  & 0\\\hline\hline
    \end{tabular} 
\end{table}

\begin{table}[htpb]
      \renewcommand{\arraystretch}{1.5}
     \setlength{\tabcolsep}{0.2cm}
     \centering
     \caption{The same as Table \ref{table:vvvv1}, but for the
     strangeness=0, isospin=0 and spin=2 channel.\label{table:vvvv3}}
     \begin{tabular}{c|ccccc}
     \hline\hline
       &  $K^*\bar{K}^*$ & $\rho\rho$ & $\omega\omega$  & $\omega\phi$  & $\phi\phi$ \\\hline
 $K^*\bar{K}^*$ &   $-3 g^2$ & $-\sqrt{3} g^2$ & $g^2$ & $-2g^2$ & $2 g^2$ \\
 $\rho\rho$          &    & $-4 g^2$ & 0 & 0 & 0 \\
 $\omega\omega$      &    &  & 0 & 0 & 0 \\
 $\omega\phi$        & & &  & 0 & 0 \\
 $\phi\phi$          &  &  &  &  & 0\\\hline\hline
    \end{tabular} 
\end{table}

\begin{table}[htpb]
      \renewcommand{\arraystretch}{1.5}
     \setlength{\tabcolsep}{0.2cm}
     \centering
     \caption{The same as Table \ref{table:vvvv1}, but for the
     strangeness=0, isospin=1 and spin=0 channel.\label{table:vvvv4}}
     \begin{tabular}{c|cccc}
     \hline\hline
       &  $K^*\bar{K}^*$ & $\rho\rho$ & $\rho\omega$  & $\rho\phi$ \\\hline
 $K^*\bar{K}^*$ &    $2 g^2$ & 0 & $-2\sqrt{2} g^2$ & $4 g^2$ \\
 $\rho\rho$          &   & 0 & 0 & 0 \\
 $\rho\omega$      &    & & 0 & 0 \\
 $\rho\phi$        & &  &  & 0 \\\hline\hline
    \end{tabular} 
\end{table}

\begin{table}[htpb]
      \renewcommand{\arraystretch}{1.5}
     \setlength{\tabcolsep}{0.4cm}
     \centering
     \caption{The same as Table \ref{table:vvvv1}, but for
     the strangeness=0, isospin=1 and spin=1 channel.\label{table:vvvv5}}
     \begin{tabular}{c|cccc}
     \hline\hline
       &  $K^*\bar{K}^*$ & $\rho\rho$ & $\rho\omega$  & $\rho\phi$ \\\hline
 $K^*\bar{K}^*$ &   $3 g^2$ & $3\sqrt{2} g^2$ & 0 & 0 \\
 $\rho\rho$          &    & $6 g^2$ & 0 & 0 \\
 $\rho\omega$      & &  & 0 & 0 \\
 $\rho\phi$        &  &  &  & 0 \\\hline\hline
    \end{tabular} 
\end{table}

\begin{table}[htpb]
      \renewcommand{\arraystretch}{1.5}
     \setlength{\tabcolsep}{0.4cm}
     \centering
     \caption{The same as Table \ref{table:vvvv1}, but for the strangeness=0, isospin=1 and
     spin=2 channel.\label{table:vvvv6}}
     \begin{tabular}{c|cccc}
     \hline\hline
       &  $K^*\bar{K}^*$ & $\rho\rho$ & $\rho\omega$  & $\rho\phi$ \\\hline
 $K^*\bar{K}^*$ &    $-g^2$ & 0 & $ \sqrt{2}g^2$ & $-2 g^2$ \\
 $\rho\rho$          &    & 0 & 0 & 0 \\
 $\rho\omega$      &  &  & 0 & 0 \\
 $\rho\phi$        & &  &  & 0 \\\hline\hline
    \end{tabular} 
\end{table}

\begin{table}[htpb]
      \renewcommand{\arraystretch}{1.5}
     \setlength{\tabcolsep}{0.2cm}
     \centering
     \caption{The same as Table \ref{table:vvvv1}, but for
     the strangeness=0, isospin=2 and spin=0(1,2) channel.\label{table:vvvv7}}
     \begin{tabular}{c|ccc}
     \hline\hline
       &  $\rho\rho$ (Spin=0) & $\rho\rho$ (Spin=1) & $\rho\rho$ (Spin=2) \\\hline
 $\rho\rho$ (Spin=0) &    $-4g^2$ & 0 & 0\\
 $\rho\rho$ (Spin=1) &     & 0 & 0  \\
 $\rho\rho$ (Spin=2) &   &  & $2g^2$  \\\hline\hline
    \end{tabular} 
\end{table}

\begin{table}[htpb]
      \renewcommand{\arraystretch}{1.5}
     \setlength{\tabcolsep}{0.4cm}
     \centering
     \caption{The same as Table \ref{table:vvvv1}, but for
     the strangeness=1, isospin=1/2 and spin=0 channel.\label{table:vvvv8}}
     \begin{tabular}{c|ccc}
     \hline\hline
       & $\rho K^*$ & $K^*\omega$ & $K^*\phi$   \\\hline
 $\rho K^*$ &   $5 g^2$ & $\sqrt{3} g^2$ & $-\sqrt{6} g^2$ \\
 $K^*\omega$          &    & $-g^2$ & $\sqrt{2}g^2$ \\
 $K^*\phi$      &  & & $-2g^2$   \\\hline\hline
    \end{tabular} 
\end{table}

\begin{table}[htpb]
      \renewcommand{\arraystretch}{1.5}
     \setlength{\tabcolsep}{0.4cm}
     \centering
     \caption{The same as Table \ref{table:vvvv1}, but
     for the strangeness=1, isospin=1/2 and spin=1 channel.\label{table:vvvv9}}
     \begin{tabular}{c|ccc}
     \hline\hline
       & $\rho K^*$ & $K^*\omega$ & $K^*\phi$   \\\hline
 $\rho K^*$ &    $\frac{9 g^2}{2}$ & $\frac{3 \sqrt{3} g^2}{2}$ & $-3 \sqrt{\frac{3}{2}} g^2$ \\
 $K^*\omega$          &    & $\frac{3 g^2}{2}$ & $-\frac{3 g^2}{\sqrt{2}}$ \\
 $K^*\phi$      &  &  &  $3 g^2$ \\\hline\hline
    \end{tabular} 
\end{table}

\begin{table}[htpb]
      \renewcommand{\arraystretch}{1.5}
     \setlength{\tabcolsep}{0.4cm}
     \centering
     \caption{The same as Table \ref{table:vvvv1}, but for the
     strangeness=1, isospin=1/2 and spin=2 channel.\label{table:vvvv10}}
     \begin{tabular}{c|ccc}
     \hline\hline
       & $\rho K^*$ & $K^*\omega$ & $K^*\phi$   \\\hline
 $\rho K^*$ &    $ -\frac{5 g^2}{2}$ & $-\frac{1}{2} \sqrt{3} g^2$ & $\sqrt{\frac{3}{2}} g^2$ \\
 $K^*\omega$          &   & $\frac{g^2}{2}$ & $-\frac{g^2}{ \sqrt{2}}$ \\
 $K^*\phi$      &   & & $g^2$ \\\hline\hline
    \end{tabular} 
\end{table}

\begin{table}[htpb]
      \renewcommand{\arraystretch}{1.5}
     \setlength{\tabcolsep}{0.1cm}
     \centering
     \caption{The same as Table \ref{table:vvvv1}, but
     for the strangeness=1, isospin=3/2 and spin=0(1,2) channel.\label{table:vvvv11}}
     \begin{tabular}{c|ccc}
     \hline\hline
       &  $\rho K^*$ (Spin=0) & $\rho K^*$ (Spin=1) & $\rho K^*$ (Spin=2) \\\hline
 $\rho K^*$ (Spin=0) &    $-4g^2$ & 0 & 0\\
 $\rho K^*$ (Spin=1) &     & 0 & 0  \\
 $\rho K^*$ (Spin=2) &   &  & $2g^2$  \\\hline\hline
    \end{tabular} 
\end{table}

\begin{table}[htpb]
      \renewcommand{\arraystretch}{1.5}
     \setlength{\tabcolsep}{0.0cm}
     \centering
     \caption{The same as Table \ref{table:vvvv1}, but for the
     strangeness=2, isospin=0 and spin=0(1,2) channel.\label{table:vvvv12}}
     \begin{tabular}{c|ccc}
     \hline\hline
       &  $K^* K^*$ (Spin=0) & $K^* K^*$ (Spin=1) & $K^* K^*$ (Spin=2) \\\hline
 $K^* K^*$ (Spin=0) &    0 & 0 & 0\\
 $K^* K^*$ (Spin=1) &     & 0 & 0  \\
 $K^* K^*$ (Spin=2) &     &  & 0  \\\hline\hline
    \end{tabular} 
\end{table}

\begin{table}[htpb]
      \renewcommand{\arraystretch}{1.5}
     \setlength{\tabcolsep}{0.0cm}
     \centering
     \caption{The same as Table \ref{table:vvvv1}, but
     for the strangeness=2, isospin=1 and spin=0(1,2) channel.\label{table:vvvv13}}
     \begin{tabular}{c|ccc}
     \hline\hline
       &  $K^* K^*$ (Spin=0) & $K^* K^*$ (Spin=1) & $K^* K^*$ (Spin=2) \\\hline
 $K^* K^*$ (Spin=0) &    $-4g^2$ & 0 & 0\\
 $K^* K^*$ (Spin=1) &     & 0 & 0  \\
 $K^* K^*$ (Spin=2) &     &  & $2g^2$  \\\hline\hline
    \end{tabular} 
 \end{table}

\begin{table*}[htpb]
      \renewcommand{\arraystretch}{1.5}
     \centering
     \caption{The $V_{ij}$'s for the $t$($u$)-channel vector-exchange diagrams in
     the strangeness=0, isospin=0 and spin=0(2) channel.\label{table:3v1}}
     \resizebox{\textwidth}{!}{%
     \begin{tabular}{c|ccccc}
     \hline\hline
       &  $K^*\bar{K}^*$ & $\rho\rho$ & $\omega\omega$  & $\omega\phi$  & $\phi\phi$ \\\hline
 $K^*\bar{K}^*$ &   $\frac{g^2 \left(M_{\rho }^2 M_{\phi }^2+\left(2 M_{\rho }^2+3 M_{\phi }^2\right) M_{\omega }^2\right) \left(4 M_{K^*}^2-3 s\right)}{4 M_{\rho }^2 M_{\phi }^2 M_{\omega }^2}$ & $\frac{\sqrt{3} g^2 \left(2 M_{\rho }^2+2 M_{K^*}^2-3 s\right)}{2 M_{K^*}^2} $ & $-\frac{g^2 \left(2 M_{\omega }^2+2 M_{K^*}^2-3 s\right)}{2  M_{K^*}^2}$ & $\frac{g^2 \left(M_{\phi }^2+M_{\omega }^2+2 M_{K^*}^2-3 s\right)}{ M_{K^*}^2}$ & $\frac{g^2 \left(-2 M_{\phi }^2-2 M_{K^*}^2+3 s\right)}{M_{K^*}^2}$ \\
 $\rho\rho$          &    & $2 g^2 \left(4-\frac{3 s}{M_{\rho }^2}\right)$ & 0 & 0 & 0 \\
 $\omega\omega$      &   & & 0 & 0 & 0 \\
 $\omega\phi$      &  & &  & 0 & 0 \\
 $\phi\phi$          &  &  &  &  & 0\\\hline\hline
    \end{tabular}} 
       \end{table*}

\begin{table*}[htpb]
      \renewcommand{\arraystretch}{1.5}
     \setlength{\tabcolsep}{0.5cm}
     \centering
     \caption{The same as Table \ref{table:3v1}, but for the
     strangeness=0, isospin=0 and spin=1 channel.\label{table:3v2}}
     \begin{tabular}{c|ccccc}
     \hline\hline
       &  $K^*\bar{K}^*$ & $\rho\rho$ & $\omega\omega$  & $\omega\phi$  & $\phi\phi$ \\\hline
       $K^*\bar{K}^*$ &  $\frac{g^2 \left(M_{\rho }^2 M_{\phi }^2+\left(2 M_{\rho }^2+3 M_{\phi }^2\right) M_{\omega }^2\right) \left(4 M_{K^*}^2-3 s\right)}{4 M_{\rho }^2 M_{\phi }^2 M_{\omega }^2}$ & 0 & 0 & 0 & 0  \\
 $\rho\rho$         & & 0 & 0 & 0 & 0 \\
 $\omega\omega$      &  &  & 0 & 0 & 0 \\
 $\omega\phi$      &  &  &  & 0 & 0 \\
 $\phi\phi$          &   &  &  &  & 0\\\hline\hline
\end{tabular}
\end{table*}

\begin{table*}[htpb]
      \renewcommand{\arraystretch}{1.5}
     \setlength{\tabcolsep}{0.4cm}
     \centering
     \caption{The same as Table \ref{table:3v1}, but for the
     strangeness=0, isospin=1 and spin=0(2) channel.\label{table:3v3}}
     \begin{tabular}{c|cccc}
     \hline\hline
       &  $K^*\bar{K}^*$ & $\rho\rho$ & $\rho\omega$  & $\rho\phi$ \\\hline
 $K^*\bar{K}^*$ &    $\frac{g^2 \left(M_{\rho }^2 M_{\phi }^2-\left(M_{\phi }^2-2 M_{\rho }^2\right) M_{\omega }^2\right) \left(4 M_{K^*}^2-3 s\right)}{4 M_{\rho }^2 M_{\phi }^2 M_{\omega }^2}$ & 0 & $-\frac{g^2 \left(M_{\rho }^2+M_{\omega }^2+2 M_{K^*}^2-3 s\right)}{ \sqrt{2} M_{K^*}^2}$ & $\frac{g^2 \left(M_{\rho }^2+M_{\phi }^2+2 M_{K^*}^2-3 s\right)}{ M_{K^*}^2}$ \\
 $\rho\rho$          &   & 0 & 0 & 0 \\
 $\rho\omega$      &      &  & 0 & 0 \\
 $\rho\phi$      &   &  &  & 0 \\\hline\hline
    \end{tabular}
       \end{table*}

\begin{table*}[htpb]
      \renewcommand{\arraystretch}{1.5}
     \setlength{\tabcolsep}{0.4cm}
     \centering
     \caption{The same as Table \ref{table:3v1}, but for
     strangeness=0, isospin=1 and spin=1 channel. \label{table:3v4}}
     \begin{tabular}{c|cccc}
     \hline\hline
       &  $K^*\bar{K}^*$ & $\rho\rho$ & $\rho\omega$  & $\rho\phi$ \\\hline
 $K^*\bar{K}^*$ &   $\frac{g^2 \left(M_{\rho }^2 M_{\phi }^2-\left(M_{\phi }^2-2 M_{\rho }^2\right) M_{\omega }^2\right) \left(4 M_{K^*}^2-3 s\right)}{4 M_{\rho }^2 M_{\phi }^2 M_{\omega }^2}$ & $\frac{g^2 \left(2 M_{\rho }^2+2 M_{K^*}^2-3 s\right)}{\sqrt{2} M_{K^*}^2}$ & 0 & 0 \\
 $\rho\rho$          &     & $g^2 \left(4-\frac{3 s}{M_{\rho }^2}\right)$ & 0 & 0 \\
 $\rho\omega$      &    &  & 0 & 0 \\
 $\rho\phi$      &    &  &  & 0 \\\hline\hline
    \end{tabular}
       \end{table*}

 \begin{table*}[htpb]
      \renewcommand{\arraystretch}{1.5}
     \setlength{\tabcolsep}{0.4cm}
     \centering
     \caption{The same as Table \ref{table:3v1}, but
     for the strangeness=0, isospin=2 and spin=0(1,2) channel.\label{table:3v5}}
     \begin{tabular}{c|ccc}
     \hline\hline
       &  $\rho\rho$ (Spin=0) & $\rho\rho$ (Spin=1) & $\rho\rho$ (Spin=2) \\\hline
 $\rho\rho$ (Spin=0) &    $g^2 \left(\frac{3 s}{M_{\rho }^2}-4\right)$ & 0 & 0\\
 $\rho\rho$ (Spin=1) &     & 0 & 0  \\
 $\rho\rho$ (Spin=2) &   &  & $g^2 \left(\frac{3 s}{M_{\rho }^2}-4\right)$  \\\hline\hline
    \end{tabular}
       \end{table*}

\begin{table*}
      \renewcommand{\arraystretch}{2.0}
     \setlength{\tabcolsep}{0.1cm}
     \centering
     \caption{The same as Table \ref{table:3v1}, but for
     strangeness=1, isospin=1/2 and spin=0(2) channel.\label{table:3v6}}

     \begin{tabular}{c|cc}
     \hline\hline
        &  $\rho K^*$ & $K^*\omega$   \\\hline
 $\rho K^*$ & $\frac{g^2 \left(4 M_{K^*}^6+\left(8 s-9 M_{\rho }^2\right) M_{K^*}^4+2 \left(3 M_{\rho }^4+5 s M_{\rho }^2-6 s^2\right) M_{K^*}^2-M_{\rho }^2 \left(M_{\rho }^4-2 s M_{\rho }^2+3 s^2\right)\right)}{4 s M_{\rho }^2 M_{K^*}^2}$ & $\frac{\sqrt{3} g^2 \left(-M_{K^*}^4+\left(M_{\rho }^2+M_{\omega }^2+2 s\right) M_{K^*}^2+\left(s-M_{\rho }^2\right) M_{\omega }^2+s \left(M_{\rho }^2-3 s\right)\right)}{4 s M_{K^*}^2}$
   \\
  $K^*\omega$ & &$\frac{g^2 \left(M_{\omega }^4-2 s M_{\omega }^2+M_{K^*}^4+3 s^2-2 \left(M_{\omega }^2+s\right) M_{K^*}^2\right)}{4 s M_{K^*}^2}$
\\\hline\hline


          & $K^*\phi$  \\\hline
 $\rho K^*$ &  $\frac{\sqrt{\frac{3}{2}} g^2 \left(M_{K^*}^4-\left(M_{\rho }^2+M_{\phi }^2+2 s\right) M_{K^*}^2+\left(M_{\rho }^2-s\right) M_{\phi }^2+s \left(3 s-M_{\rho }^2\right)\right)}{2 s M_{K^*}^2}$
   \\
   $K^*\omega$  & $ \frac{g^2 \left(-M_{K^*}^4+\left(M_{\phi }^2+M_{\omega }^2+2 s\right) M_{K^*}^2+\left(s-M_{\phi }^2\right) M_{\omega }^2+s \left(M_{\phi }^2-3 s\right)\right)}{2 \sqrt{2} s M_{K^*}^2}$ \\
 $K^*\phi$   & $\frac{g^2 \left(M_{\phi }^4-2 s M_{\phi }^2+M_{K^*}^4+3 s^2-2 \left(M_{\phi }^2+s\right) M_{K^*}^2\right)}{2 s M_{K^*}^2}$
\\\hline\hline
    \end{tabular}

\end{table*}

\begin{table*}[htpb]
      \renewcommand{\arraystretch}{2.0}
     \setlength{\tabcolsep}{0.0cm}
     \centering
     \caption{The same as Table \ref{table:3v1}, but
     for the strangeness=1, isospin=1/2 and spin=1 channel.\label{table:3v7}}
     \begin{tabular}{c|cc}
     \hline\hline
       &  $\rho K^*$ & $K^*\omega$   \\\hline
 $\rho K^*$& $\frac{g^2 \left(M_{\rho }^6-2 s M_{\rho }^4+3 s^2 M_{\rho }^2+4 M_{K^*}^6+\left(8 s-7 M_{\rho }^2\right) M_{K^*}^4+2 \left(M_{\rho }^4+3 s M_{\rho }^2-6 s^2\right) M_{K^*}^2\right)}{4 s M_{\rho }^2 M_{K^*}^2} $
 & $\frac{\sqrt{3} g^2 \left(-M_{K^*}^4+\left(M_{\rho }^2+M_{\omega }^2+2 s\right) M_{K^*}^2+\left(s-M_{\rho }^2\right) M_{\omega }^2+s \left(M_{\rho }^2-3 s\right)\right)}{4 s M_{K^*}^2}$ \\

 $K^*\omega$ & &$-\frac{g^2 \left(M_{\omega }^4-2 s M_{\omega }^2+M_{K^*}^4+3 s^2-2 \left(M_{\omega }^2+s\right) M_{K^*}^2\right)}{4 s M_{K^*}^2} $
 \\\hline\hline


       & $K^*\phi$  \\\hline
 $\rho K^*$
 &  $\frac{\sqrt{\frac{3}{2}} g^2 \left(M_{K^*}^4-\left(M_{\rho }^2+M_{\phi }^2+2 s\right) M_{K^*}^2+\left(M_{\rho }^2-s\right) M_{\phi }^2+s \left(3 s-M_{\rho }^2\right)\right)}{2 s M_{K^*}^2}$ \\
  $K^*\omega$  &$\frac{g^2 \left(M_{K^*}^4-\left(M_{\phi }^2+M_{\omega }^2+2 s\right) M_{K^*}^2+\left(M_{\phi }^2-s\right) M_{\omega }^2+s \left(3 s-M_{\phi }^2\right)\right)}{2 \sqrt{2} s M_{K^*}^2} $\\

 $K^*\phi$   &$-\frac{g^2 \left(M_{\phi }^4-2 s M_{\phi }^2+M_{K^*}^4+3 s^2-2 \left(M_{\phi }^2+s\right) M_{K^*}^2\right)}{2 s M_{K^*}^2}$
  \\\hline\hline
  \end{tabular}
\end{table*}

\begin{table*}[htpb]
      \renewcommand{\arraystretch}{2.0}
     \setlength{\tabcolsep}{0.4cm}
     \centering
     \caption{The same as Table \ref{table:3v1}, but
     for the strangeness=1, isospin=3/2 and spin=0(2) and Spin=1 channel. \label{table:3v8}}
     \begin{tabular}{c|c}
     \hline\hline
 $\rho K^*$ [Spin=0(2)] &    $\frac{g^2 \left(M_{\rho }^6-2 s M_{\rho }^4+3 s^2 M_{\rho }^2-M_{K^*}^6+\left(3 M_{\rho }^2-2 s\right) M_{K^*}^4+\left(-3 M_{\rho }^4-4 s M_{\rho }^2+3 s^2\right) M_{K^*}^2\right)}{2 s M_{\rho }^2 M_{K^*}^2}$  \\
 $\rho K^*$ (Spin=1)     & $-\frac{g^2 \left(M_{K^*}^2-M_{\rho }^2\right) \left(-M_{\rho }^4+2 s M_{\rho }^2+M_{K^*}^4-3 s^2+2 s M_{K^*}^2\right)}{2 s M_{\rho }^2 M_{K^*}^2}$
   \\\hline\hline
    \end{tabular}
\end{table*}

\begin{table*}[htpb]
      \renewcommand{\arraystretch}{1.5}
     \setlength{\tabcolsep}{0.4cm}
     \centering
     \caption{The same as Table \ref{table:3v1}, but for the
     strangeness=2, isospin=0 and spin=0(1,2) channel.\label{table:3v9}}
     \begin{tabular}{c|ccc}
     \hline\hline
     & $K^* K^*$ (Spin=0)& $ K^* K^*$ (Spin=1) & $K^* K^*$ (Spin=2)\\\hline
         $K^* K^*$ (Spin=0) & 0 & 0 &0\\
 $K^* K^*$ (Spin=1) &     & $\frac{g^2 \left(\left(3 M_{\phi }^2-2 M_{\rho }^2\right) M_{\omega }^2-M_{\rho }^2 M_{\phi }^2\right) \left(4 M_{K^*}^2-3 s\right)}{4 M_{\rho }^2 M_{\phi }^2 M_{\omega }^2}$ &0 \\
 $K^* K^*$ (Spin=2) &  & &0   \\\hline\hline
    \end{tabular}
       \end{table*}

\begin{table*}[htpb]
      \renewcommand{\arraystretch}{2.0}
     \setlength{\tabcolsep}{0.1cm}
     \centering
     \caption{The same as Table \ref{table:3v1}, but for the
     strangeness=2, isospin=1 and spin=0(1,2) channel. \label{table:3v10}}
     \begin{tabular}{c|ccc}
     \hline\hline
     & $K^* K^*$ (Spin=0)& $ K^* K^*$ (Spin=1) & $K^* K^*$ (Spin=2)\\\hline
         $K^* K^*$ (Spin=0) & $-\frac{g^2 \left(M_{\rho }^2 M_{\phi }^2+\left(2 M_{\rho }^2+M_{\phi }^2\right) M_{\omega }^2\right) \left(4 M_{K^*}^2-3 s\right)}{4 M_{\rho }^2 M_{\phi }^2 M_{\omega }^2}$ & 0 &0\\
 $K^* K^*$ (Spin=1) &     & 0 &0 \\
 $K^* K^*$ (Spin=2) &  & & $-\frac{g^2 \left(M_{\rho }^2 M_{\phi }^2+\left(2 M_{\rho }^2+M_{\phi }^2\right) M_{\omega }^2\right) \left(4 M_{K^*}^2-3 s\right)}{4 M_{\rho }^2 M_{\phi }^2 M_{\omega }^2}$   \\\hline\hline
    \end{tabular}
       \end{table*}

\clearpage
\subsection{Box diagram amplitudes}
In this section, we provide the explicit box diagram amplitudes, corresponding
to Eq.~(22), for
different strangeness and isospin  but only spin=0 channels.
Those amplitudes for spin=2 channels can be obtained by multiplying $2/5$ to
the corresponding spin=0 amplitudes, as explained in the main text.
To simplify the expressions, we have used the abbreviations defined in Table \ref{table:abb}.
T

\begin{table}[t]
     \setlength{\tabcolsep}{0.2cm}
     \centering
     \caption{The abbreviations used in calculating the box diagrams:   $\tilde{G}_i=G_4(m_{p1},m_{p2},m_{p3},m_{p4},s,k^0_1,k^2_0,k_3^0,k_4^0)$ with $i=1\cdots 20$ and
$p1$, $p2$, $p3$, $p4$ the particles appearing in the box diagram with the order as given in
Fig.~2.
In the text, $\tilde{G}_i(u)=G_4(m_{p1},m_{p2},m_{p3},m_{p4},s,k^0_1,k^2_0,k_4^0,k_3^0)$.\label{table:abb}}
     \begin{tabular}{c|cccc||c|cccc}
     \hline\hline
 $i$ & $p1$  & $p2$  & $p3$  & $p4$ &
 $i$ & $p1$  & $p2$  & $p3$  & $p4$ \\\hline
 1   & $\eta$ & $K$    & $\eta$ & $K$ &
 2   & $\eta$ & $K$    & $\pi$  & $K$ \\
 3   & $K$    & $\eta$ &  $K$   & $\eta$ &
 4   & $K$    & $\pi$  & $K$    & $\pi$ \\
 5   & $\pi$  & $K$    & $\eta$ & $K$ &
 6   & $\pi$  & $K$    & $\pi$  & $K$ \\
 7   & $\eta$ & $K$    & $K$    & $K$ &
 8   & $K$    & $\pi$  & $\pi$  & $\pi$\\
 9   & $\pi$  & $K$    & $K$    & $K$ &
 10   & $K$    & $K$    & $K$    & $K$\\
 11   & $\pi$  & $\pi$  & $\pi$  & $\pi$&
 12   & $K$    & $\eta$ & $K$    & $\pi$\\
 13   & $K$    & $\pi$  & $K$    & $\eta$&
 14   & $K$    & $K$    & $K$    & $\eta$\\
 15   & $K$    & $K$    & $K$    & $\pi$&
 16   & $K$    & $K$    & $\pi$  & $\pi$\\
 17   & $\pi$  & $\pi$  & $K$    & $K$&
 18   & $\pi$  & $\pi$  & $\pi$  & $K$\\
 19   & $K$    & $\eta$ & $K$    & $K$&
 20   & $K$    & $\pi$  & $K$    & $K$\\
\hline\hline
    \end{tabular}
  \end{table}

\begin{enumerate}
\item Strangeness=0, isospin=0, and spin=0:
 There are five channels, i.e., $K^*\bar{K}^*$, $\rho\rho$, $\omega\omega$, $\omega\phi$, $\phi\phi$, with the order of 1, 2, 3, 4, 5:
       \begin{eqnarray*}\
        v_{1,1}&=&60 g^4(3 \tilde{G}_{1} + 3 \tilde{G}_{2} + 12 \tilde{G}_3 + 4 \tilde{G}_4 + 3  \tilde{G}_5 + 3 \tilde{G}_6),\\
    v_{1,2}&=&40 \sqrt{3} g^4 (3 \tilde{G}_7 +8 \tilde{G}_8 +3 \tilde{G}_9),\\
    v_{1,3}&=&-120 g^4 (\tilde{G}_7+\tilde{G}_9),\\
    v_{1,4}&=&120 g^4 (\tilde{G}_7 + \tilde{G}_7(u) +\tilde{G}_9 +\tilde{G}_9(u))\\
     v_{1,5}&=&-240 g^4 (\tilde{G}_7 + \tilde{G}_9)\\
     v_{2,2}&=&80 g^4 (3 \tilde{G}_{10} +16 \tilde{G}_{11})\\
     v_{2,3}&=&-80 \sqrt{3} g^4 \tilde{G}_{10} \\
     v_{2,4}&=&80 \sqrt{3} g^4 (\tilde{G}_{10} +\tilde{G}_{10}(u)),\\
     v_{2,5}&=&-160 \sqrt{3} g^4 \tilde{G}_{10},\\
     v_{3,3}&=&80 g^4 \tilde{G}_{10},\\
     v_{3,4}&=&-80 g^4 (\tilde{G}_{10}+\tilde{G}_{10}(u)),\\
      v_{3,5}&=&160 g^4 \tilde{G}_{10},\\
      v_{4,4}&=&160 g^4 (\tilde{G}_{10}+\tilde{G}_{10}(u)),\\
       v_{4,5}&=&-320 g^4 \tilde{G}_{10},\\
       v_{5,5}&=&320 g^4 \tilde{G}_{10}.
     \end{eqnarray*}
\item Strangeness=0, isospin=1, and spin=0:
There are four channels, i.e.,  $K^*\bar{K}^*$, $\rho\rho$, $\rho\omega$, $\rho\phi$,
with the order 1, 2, 3, 4:
     \begin{eqnarray*}
     v_{1,1}&=&20 g^4 (9 \tilde{G}_{1} -3\tilde{G}_{2} +12 \tilde{G}_{12}+ 12 \tilde{G}_{13} -3\tilde{G}_5 + \tilde{G}_6)\\
      v_{1,2}&=&0,\\
         v_{13}&=&-20 \sqrt{2} g^4 (3 \tilde{G}_7 +3 \tilde{G}_7(u) - \tilde{G}_9 - \tilde{G}_9(u))\\
     v_{1,4}&=&40 g^4 (3 \tilde{G}_7 +3 \tilde{G}_7(u) - \tilde{G}_9 - \tilde{G}_9(u))\\
       v_{2,2}&=&v_{2,3}=v_{2,4}=0,\\
          v_{3,3}&=& 80 g^4 (\tilde{G}_{10} + \tilde{G}_{10}(u))),\\
     v_{3,4}&=&-80 \sqrt{2}g^4 (\tilde{G}_{10} + \tilde{G}_{10}(u))\\
     v_{4,4}&=&160 g^4 (\tilde{G}_{10} + \tilde{G}_{10}(u))
     \end{eqnarray*}

\item Strangeness=0, isospin=2, and spin=0:
There is only one channel in this sector, i.e., $\rho\rho$:
     \begin{eqnarray*}
     v=320 g^4 \tilde{G}_{11}
     \end{eqnarray*}

\item Strangeness=1, isospin=1/2, and spin=0:
There are three channels, i.e.,  $\rho K^*$, $K^*\omega$, and $K^*\phi$, with the
order 1, 2, 3:
         \begin{eqnarray*}
     v_{1,1}&=&20 g^4 ( 9 \tilde{G}_{14} + \tilde{G}_{15} + 4 \tilde{G}_{16}(u) + 4 \tilde{G}_{17}(u) + 16 \tilde{G}_{18})\\
     v_{1,2}&=&-20 \sqrt{3} g^4 (3 \tilde{G}_{14}(u)-\tilde{G}_{15}(u)-4 \tilde{G}_{17} )\\
     v_{1,3}&=&20 \sqrt{6} g^4 (3 \tilde{G}_{14}(u)-\tilde{G}_{15}(u)-4 \tilde{G}_{17})\\
     v_{2,2}&=&60 g^4 (\tilde{G}_{19}+\tilde{G}_{20}),\\
     v_{2,3}&=&-60 \sqrt{2} g^4 (\tilde{G}_{19}+\tilde{G}_{20})\\
     v_{3,3}&=&120 g^4 (\tilde{G}_{19}+\tilde{G}_{20})
     \end{eqnarray*}

\item Strangeness=1, isospin=1/2, and spin=0:
There is only one channel in this sector, i.e., $\rho K^*$:
     \begin{eqnarray*}
     v&=&80 g^4 (\tilde{G}_{15}+\tilde{G}_{16}(u)+\tilde{G}_{17}(u)+\tilde{G}_{18})
     \end{eqnarray*}

\item Strangeness=2, isospin=0, and spin=0:
There is only one channel in this sector, i.e., $K^* K^*$:
         \begin{eqnarray*}
     v=0.
         \end{eqnarray*}
\item Strangeness=2, isospin=1, and spin=0:
There is only one channel in this sector, i.e., $K^* K^*$:
         \begin{eqnarray*}
      v&=&20 g^4 (9 \tilde{G}_{1}+ 3 \tilde{G}_{2}+3 \tilde{G}_5 + \tilde{G}_6)
         \end{eqnarray*}

\end{enumerate}

\subsection{$G_n$ in the evaluation of the four-point loop function}
 Here we provide the explicit form of $G_n$, which appears in
 the evaluation of the four-point loop function $G_4$ [Eq.~(\ref{eq:4ploop})]. The symbols are
 the same as in the main text, except here we have replaced $k^0_1$, $k^0_2$,
 $k^0_3$, and $k_4^0$ by $E_1$, $E_2$, $E_3$ and $E_4$.
\begin{widetext}
\begin{eqnarray}
G_\mathrm{n}&=&
\omega _1 \left(\omega _3 \left(\omega _2+\omega _4\right) E_3^2-2 P^0 \omega _3 \omega _4 E_3-\left(\omega _2+\omega _3\right) \left(\left(\omega _3+\omega _4\right) \omega _2^2
+\left(\omega _3^2+3 \omega _4 \omega _3+2 \omega _4^2\right) \omega _2+\omega _4 \left(\left(\omega _3+\omega _4\right)^2\right.\right.\right.\nonumber\\
&&\left.\left.\left.-s\right)\right)\right) E_1^4\nonumber\\
&+&2 \omega _1 \left(-\omega _3 \left(\omega _2+\omega _4\right) E_3^3+P^0 \omega _3 \omega _4 E_3^2+\omega _3 \left(\omega _2^3+2 \left(\omega _3+\omega _4\right) \omega _2^2+\left(\omega _3^2+4 \omega _4 \omega _3+2 \omega _4^2\right) \omega _2+\right.\right.\nonumber\\
&&\left.\left. \omega _4 \left(s+\left(\omega _3+\omega _4\right)^2\right)\right) E_3+P^0 \left(\omega _2+\omega _3\right) \omega _4 \left(-s+\omega _2^2+\left(\omega _3+\omega _4\right){}^2+\omega _2 \left(\omega _3+2 \omega _4\right)\right)\right) E_1^3\nonumber\\
&+&\omega _1 \left(\omega _3 \left(\omega _2+\omega _4\right) E_3^4+2 P^0 \omega _3 \omega _4 E_3^3-2 \omega _3 \left(\omega _2^3+\omega _3 \omega _2^2+\omega _3 \left(\omega _3+2 \omega _4\right) \omega _2+\omega _1^2 \left(\omega _2+\omega _4\right)+\right.\right.\nonumber\\
&&\left.\left. \omega _1 \left(\omega _2+\omega _4\right) \left(\omega _2+\omega _3+\omega _4\right)+\omega _4 \left(3 s+\omega _3^2+\omega _4^2+\omega _3 \omega _4\right)\right) E_3^2+\right.\nonumber\\
&&\left. 2 P^0 \omega _3 \omega _4 \left(s+2 \omega _1^2-2 \omega _2^2-\omega _3^2-\omega _4^2-4 \omega _2 \omega _3-4 \omega _2 \omega _4-4 \omega _3 \omega _4+2 \omega _1 \left(\omega _2+\omega _3+\omega _4\right)\right) E_3+\right.\nonumber\\
&&\left. \left(\omega _2+\omega _3\right) \left(\left(\omega _3+\omega _4\right) \omega _2^4+\left(\omega _3^2+3 \omega _4 \omega _3+2 \omega _4^2\right) \omega _2^3+\left(\omega _3^3+5 \omega _4 \omega _3^2+6 \omega _4^2 \omega _3+2 \omega _4^3-2 s \omega _4\right) \omega _2^2+\right.\right.\nonumber\\
&&\left.\left. \left(\omega _3^4+3 \omega _4 \omega _3^3+6 \omega _4^2 \omega _3^2+2 \omega _4 \left(s+3 \omega _4^2\right) \omega _3+2 \omega _4^4-2 s \omega _4^2\right) \omega _2+\omega _4 \left(s^2-2 \left(\omega _3^2+\omega _4 \omega _3+\omega _4^2\right) s+\right.\right.\right.\nonumber\\
&&\left.\left.\left. \left(\omega _3+\omega _4\right)^2 \left(\omega _3^2+\omega _4^2\right)\right)+2 \omega _1^2 \left(\left(\omega _3+\omega _4\right) \omega _2^2+\left(\omega _3^2+3 \omega _4 \omega _3+2 \omega _4^2\right) \omega _2+\omega _4 \left(\left(\omega _3+\omega _4\right)^2-s\right)\right)+\right.\right.\nonumber\\
&&\left.\left.2 \omega _1 \left(\omega _2+\omega _3+\omega _4\right) \left(\left(\omega _3+\omega _4\right) \omega _2^2+\left(\omega _3^2+3 \omega _4 \omega _3+2 \omega _4^2\right) \omega _2+\omega _4 \left(\left(\omega _3+\omega _4\right)^2-s\right)\right)\right)\right) E_1^2\nonumber\\
&-&2 \omega _1 \left(P^0 \omega _3 \omega _4 E_3^4-\omega _3 \left(\omega _2^3+2 \omega _4 \omega _2^2+2 \omega _4^2 \omega _2+\omega _4^3+2 \omega _1 \left(\omega _2+\omega _4\right)^2+s \omega _4+\omega _1^2 \left(\omega _2+\omega _4\right)\right) E_3^3+\right.\nonumber\\
&&\left. P^0 \omega _3 \omega _4 \left(-s+\omega _1^2+2 \omega _2^2-2 \omega _3^2+\omega _4^2-2 \omega _2 \omega _3+4 \omega _2 \omega _4-2 \omega _3 \omega _4+\omega _1 \left(4 \omega _2-2 \omega _3+4 \omega _4\right)\right) E_3^2+\right.\nonumber\\
&&\left. \omega _3 \left(\omega _2^5+2 \left(\omega _3+\omega _4\right) \omega _2^4+\left(\omega _3^2+4 \omega _4 \omega _3+2 \omega _4^2\right) \omega _2^3+2 \omega _4 \left(s+\left(\omega _3+\omega _4\right)^2\right) \omega _2^2+\right.\right.\nonumber\\
&&\left.\left. 2 \omega _4 \left(\left(2 \omega _3-\omega _4\right) s+\omega _4 \left(\omega _3+\omega _4\right)^2\right) \omega _2+\omega _4 \left(s^2+\left(\omega _3^2-2 \omega _4 \omega _3-2 \omega _4^2\right) s+\omega _4^2 \left(\omega _3+\omega _4\right)^2\right)+\right.\right.\nonumber\\
&&\left.\left. \omega _1^2 \left(\omega _2^3+2 \left(\omega _3+\omega _4\right) \omega _2^2+\left(\omega _3^2+4 \omega _4 \omega _3+2 \omega _4^2\right) \omega _2+\omega _4 \left(s+\left(\omega _3+\omega _4\right)^2\right)\right)+\right.\right.\nonumber\\
&&\left.\left. 2 \omega _1 \left(\omega _2^4+2 \left(\omega _3+\omega _4\right) \omega _2^3+\left(\omega _3^2+4 \omega _4 \omega _3+2 \omega _4^2\right) \omega _2^2+2 \omega _4 \left(s+\left(\omega _3+\omega _4\right)^2\right) \omega _2+\right.\right.\right.\nonumber\\
&&\left.\left.\left. \omega _4 \left(\left(2 \omega _3-\omega _4\right) s+\omega _4 \left(\omega _3+\omega _4\right){}^2\right)\right)\right) E_3+\right.\nonumber\\
&&\left. P^0 \left(\omega _2+\omega _3\right) \omega _4 \left(\omega _2^4+\omega _3 \omega _2^3+2 \omega _4 \omega _2^3+\omega _3^2 \omega _2^2+\omega _4^2 \omega _2^2+2 \omega _3 \omega _4 \omega _2^2+\omega _3^3 \omega _2+\omega _3 \omega _4^2 \omega _2+2 \omega _3^2 \omega _4 \omega _2+\omega _3^4+\omega _3^2 \omega _4^2-\right.\right.\nonumber\\
&&\left.\left. s \left(\omega _1^2+2 \left(\omega _2+\omega _3\right) \omega _1+\omega _2^2+\omega _3^2+\omega _2 \omega _3\right)+2 \omega _3^3 \omega _4+\omega _1^2 \left(\omega _2^2+\left(\omega _3+2 \omega _4\right) \omega _2+\left(\omega _3+\omega _4\right)^2\right)+\right.\right.\nonumber\\
&&\left.\left. 2 \omega _1 \left(\omega _2^3+\left(\omega _3+2 \omega _4\right) \omega _2^2+\left(\omega _3+\omega _4\right)^2 \omega _2+\omega _3 (\omega _3+\omega _4)^2\right)\right)\right) E_1\nonumber\\
&-&(\omega _1+\omega _2) \left(\omega _3 \left((\omega _2+\omega _4) \omega _1^2+(\omega _2^2+3 \omega _4 \omega _2+2 \omega _4^2) \omega _1+\omega _4 \left((\omega _2+\omega _4)^2-s\right)\right) E_3^4+\right.\nonumber\\
&&\left. 2 P^0 \omega _3 \omega _4 \left(s-\omega _1^2-(\omega _2+\omega _4)^2-\omega _1 (\omega _2+2 \omega _4)\right) E_3^3-\omega _3 \left((\omega _2+\omega _4) \omega _1^4+(\omega _2+\omega _4) \left(\omega _2+2 \left(\omega _3+\omega _4\right)\right) \omega _1^3+\right.\right.\nonumber\\
&&\left.\left. \left(\omega _2^3+\left(4 \omega _3+5 \omega _4\right) \omega _2^2+2 \left(\omega _3^2+5 \omega _4 \omega _3+3 \omega _4^2\right) \omega _2+2 \omega _4 \left(-s+\omega _3^2+\omega _4^2+3 \omega _3 \omega _4\right)\right) \omega _1^2+\left(\omega _2^4+\left(2 \omega _3+3 \omega _4\right) \omega _2^3+\right.\right.\right.\nonumber\\
&&\left.\left.\left. 2 \left(\omega _3^2+5 \omega _4 \omega _3+3 \omega _4^2\right) \omega _2^2+2 \omega _4 \left(s+3 \omega _3^2+3 \omega _4^2+7 \omega _3 \omega _4\right) \omega _2+2 \omega _4 (\omega _3+\omega _4) \left(\omega _4 \left(2 \omega _3+\omega _4\right)-s\right)\right) \omega _1+\right.\right.\nonumber\\
&&\left.\left.\omega _4 \left(-s+\omega _2^2+2 \omega _3^2+\omega _4^2+2 \omega _2 \omega _3+2 \omega _3 \omega _4\right) \left(\left(\omega _2+\omega _4\right)^2-s\right)\right) E_3^2+2 P^0 \omega _3 \omega _4 \left(\omega _1^4+\left(\omega _2+2 \left(\omega _3+\omega _4\right)\right) \omega _1^3+\right.\right.\nonumber\\
&&\left.\left.\left(\omega _2^2+2 \left(\omega _3+\omega _4\right) \omega _2+\omega _3^2+\omega _4^2+4 \omega _3 \omega _4\right) \omega _1^2+\left(\omega _2^3+2 \left(\omega _3+\omega _4\right) \omega _2^2+\left(\omega _3^2+4 \omega _4 \omega _3+\omega _4^2\right) \omega _2+\right.\right.\right.\nonumber\\
&&\left.\left.\left. 2 \omega _3 \omega _4 (\omega _3+\omega _4)\right) \omega _1+(\omega _2+\omega _3)^2 (\omega _2+\omega _4)^2-s \left(\omega _1^2+\left(\omega _2+2 \omega _3\right) \omega _1+\left(\omega _2+\omega _3\right)^2\right)\right) E_3+\right.\nonumber\\
&&\left. (\omega _1+\omega _3) (\omega _2+\omega _3) \left(\left(\left(\omega _3+\omega _4\right) \omega _2^2+\left(\omega _3^2+3 \omega _4 \omega _3+2 \omega _4^2\right) \omega _2+\omega _4 \left(\left(\omega _3+\omega _4\right)^2-s\right)\right) \omega _1^3+\right.\right.\nonumber\\
&&\left.\left. \left(\omega _2+\omega _3+2 \omega _4\right) \left(\left(\omega _3+\omega _4\right) \omega _2^2+\left(\omega _3^2+3 \omega _4 \omega _3+2 \omega _4^2\right) \omega _2+\omega _4 \left(\left(\omega _3+\omega _4\right){}^2-s\right)\right) \omega _1^2+\right.\right.\nonumber\\
&&\left.\left. \left(\left(\omega _3^2+3 \omega _4 \omega _3+2 \omega _4^2\right) \omega _2^3+\left(\omega _3^3+6 \omega _4 \omega _3^2+10 \omega _4^2 \omega _3+5 \omega _4^3-s \omega _4\right) \omega _2^2+\omega _4 \left(3 \omega _3+4 \omega _4\right) \left(\left(\omega _3+\omega _4\right)^2-s\right) \omega _2+\right.\right.\right.\nonumber\\
&&\left.\left.\left.\omega _4 \left(s^2-\left(\omega _3^2+4 \omega _4 \omega _3+2 \omega _4^2\right) s+\omega _4 \left(\omega _3+\omega _4\right)^2 \left(2 \omega _3+\omega _4\right)\right)\right) \omega _1+
\right.\right.\nonumber\\
&&\left.\left. \left(\omega _2+\omega _3\right) \omega _4 \left(\left(\omega _2+\omega _4\right)^2-s\right) \left(\left(\omega _3+\omega _4\right)^2-s\right)\right)\right).
\end{eqnarray}
\end{widetext}

\end{document}